\documentclass[authoryear]{elsarticle}
\biboptions{semicolon}
\journal{Icarus}
\usepackage{natbib}
\usepackage{graphicx}
\usepackage{booktabs}
\usepackage{array}
\usepackage{rotating}
\usepackage{float}
\usepackage[caption=false]{subfig}
\usepackage{subfig}
\usepackage{amssymb,amsmath}

\def \micron {$\mu$m}

\def \Ftwo {Fe$^{2+}$}
\def \Ctwo {Ca$^{2+}$}

\begin{document}

\begin{frontmatter}
%Title and Author Block
%%%%%%%%%%%%%%%%%%%%%%%%%%%%%%%%%%%%%%%%%%%%%%%%%%%%%%%%%%%%%%%%%%%%%%%%%%%%%%%%
\title{Composition, Mineralogy, and Porosity of Multiple Asteroid Systems from Visible\tnoteref{t1} and Near-infrared Spectral Data}

\tnotetext[t1]{Based on observations obtained at the Southern Astrophysical Research (SOAR) telescope, which is a joint project of the MinistŽrio da Cincia, Tecnologia, e Inova‹o (MCTI) da Repœblica Federativa do Brasil, the U.S. National Optical Astronomy Observatory (NOAO), the University of North Carolina at Chapel Hill (UNC), and Michigan State University (MSU).}

\author[utk]{S. S. Lindsay}\fnref{fn1}
\ead{slindsay@utk.edu}

\author[seti]{F. Marchis \fnref{fn2}}
\ead{fmarchis@seti.org}

\author[utk]{J. P. Emery \fnref{fn2}}
\ead{jemery2@utk.edu}

\author[seti,run]{J. E. Enriquez \fnref{fn2}}
\ead{e.enriquez@astro.ru.nl}

\author[odv]{M. Assafin}
\ead{massaf@astro.ufrj.br}

\fntext[fn1]{Now at Atmospheric, Oceanic, and Planetary Physics, University of Oxford. E-mail: Sean.Lindsay@physics.ox.ac.uk}
\fntext[fn2]{Visiting Astronomer at the Infrared Telescope Facility, which is operated by the University of Hawaii under Cooperative Agreement no. NNX-08AE38A with the National Aeronautics and Space Administration, Science Mission Directorate, Planetary Astronomy Program.}

%Affiliations
\address[utk]{Department of Earth and Planetary Sciences, University of Tennessee, 1412 Circle Drive, Knoxville, TN 37996, USA}
\address[seti]{Carl Sagan Center, SETI Institute. 189 Bernardo Ave., Mountain View, CA 94043, USA}
\address[run]{Department of Astrophysics, Radboud University Nijmegen, PO Box 9010, 6500 GL, Nijmegen, The Netherlands}
\address[odv]{Observatorio do Valongo, UFRJ, Ladeira Pedro Antionio 43, Rio de Janeiro, Brazil}

%%%%%%%%%%%%%%%%%%%%%%%%%%%%%%%%%%%%%%%%%%%%%%%%%%%%%%%%%%%%%%%%%%%%%%%%%%%%%%%%
%ABSTRACT
\begin{abstract}
We aim to provide a taxonomic and compositional characterization of Multiple Asteroid Systems (MASs) located in the main belt (MB) using visible (0.45~--~0.85~\micron) and near-infrared (0.7~--~2.5~\micron) spectral data of 42 MB MASs.  The compositional and mineralogical analysis is applied to determine meteorite analogs for the MASs, which, in turn, are applied to the MAS density measurements of \citet{Marchis:2012} to estimate the porosity of the systems.  The macroporosities are used to evaluate the primary MAS formation hypotheses.  Our spectral survey consists of visible and near-infrared spectral data.  The visible observing campaign includes 25 MASs obtained using the Southern Astrophysical Research (SOAR) telescope with the Goodman High Throughput Spectrometer.  The infrared observing campaign includes 34 MASs obtained using the NASA Infrared Telescope Facility (IRTF) with the SpeX spectragraph.  For completeness, both visible and NIR data sets are supplemented with publicly available data, and the data sets are combined where possible.  The MASs are classified using the Bus-DeMeo taxonomic system.  In order to determine mineralogy and meteorite analog, we perform a NIR spectral band parameter analysis using a new analysis routine, the Spectral Analysis Routine for Asteroids (SARA).  The SARA routine determines band centers, areas, and depths by utilizing the diagnostic absorption features near 1- and 2-\micron  \ due to \Ftwo crystal field transitions in olivine + pyroxene and pyroxene, respectively.  The band parameter analysis provides the Gaffey subtype for the S-complex MASs; the relative abundance olivine-to-pyroxene ratio; and olivine and pyroxene modal abundances for S-complex and V-type MASs.  This mineralogical information is then applied to determine meteorite analogs.  Through applying calibration studies, we are able to determine the H, L, and LL meteorite analogs for 15 MASs with ordinary chondrite-like (OC) mineralogies.  We observe an excess (10/15) of LL-like mineralogies.  Of the ten MASs with LL-like mineralogies, seven are consistent with Flora family membership, supporting the hypothesis that the Flora family is a source of LL-like NEAs and LL chondrites on Earth.  Our band parameter analysis is unable to clearly distinguish between the HED subgroups for the 6 V-type MASs.  Using the measured densities of the meteorite analog and the MAS densities from \citet{Marchis:2012}, we estimate the macroporosity for 13 MASs.  We find that all of the MASs with estimated macroporosities are in agreement with formation hypotheses. 
\end{abstract}
\end{frontmatter}

%%%%%%%%%%%%%%%%%%%%%%%%%%%%%%%%%%%%%%%%%%%%%%%%%%%%
\section{Introduction}
\label{sec:intro}

The discovery of 243 Ida's small satellite, Dactyl, during the \emph{Galileo} fly-by launched the era of investigating asteroid systems with multiple components, or multiple asteroid systems \citep[MASs;][]{Chapman:1995,Belton:1996}.  The Ida-Dactyl system was the first detection of an asteroid with a satellite, and in the subsequent two decades since the system's discovery more than 225 MASs have been discovered.  The MASs discovered to date exhibit  a striking diversity in their key characteristics: separation of components, size ratio of components, mutual orbits, locations of the system, and system angular momentum.  For example, the Ida-Dactyl system is a large asteroid with a small companion ($D_p/D_s = 22.4,$ where $D_p$ and $D_s$ are the diameter of primary and satellite, respectively) with a small system semi-major axis that is only 3.44~$D_p$ \citep{Belton:1994,Belton:1995}, while 90 Antiope is a large (D$_p = 87.80$, \citealp{Johnston:2013}) binary asteroid system with nearly equal size components ($D_p/D_s = 1.05$) separated by 117 km \citep{Merline:2000, Descamps:2007}.  In contrast to these larger systems, 1509 Esclangona is thought to be composed of two small asteroids ($D_s/D_p = 0.5$) with $D_p = 8.0 \pm 0.8$ km and $D_s = 4 \pm 0.7$ km \citep{Marchis:2012}.  Such a wide diversity between the MASs strongly indicates that there are several mechanisms involved that govern the formation and evolution of MASs.  However, our understanding of the specific mechanisms, and the physics that govern them, remains limited.  In this study, we aim to provide context and constraints on MAS formation and evolution mechanisms through a detailed visible and near-infrared (NIR) spectral investigation to determine MAS taxonomies, meteorite analogs, and mineralogy.

\begin{figure}[!h]
	\begin{center}
		\includegraphics[height=6.0in,width=5.2in]{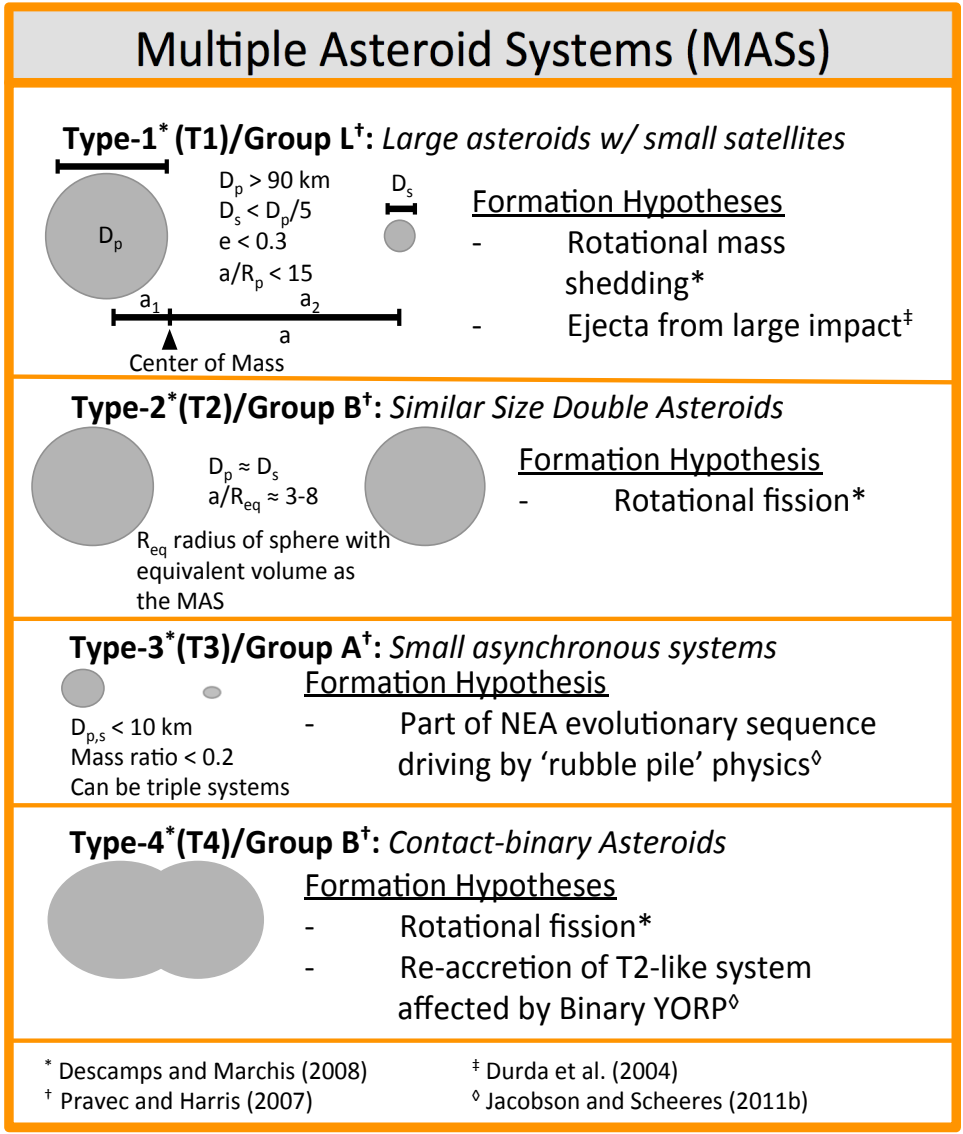}
		\caption{Graphic summarizing formation hypotheses and characteristics for the MAS types.}
		\label{fig:mas_types}
	\end{center}
\end{figure}

%\subsection{Multiple Asteroid Systems}
Previous studies separate MASs into four general categories based on their known physical parameters \citep{Descamps:2008,Pravec:2007}.  We adopt the nomenclature of \citet{Descamps:2008}, which divides the MASs into types based on the physical and orbital parameters;  these four MAS types are summarized in Fig.~\ref{fig:mas_types}.  Each of the MAS types has its own collection of hypothesized formation and evolution scenarios that are linked to the impact history and internal strength of the progenitor.
%to the internal strength of the progenitor, tidal disruption, impact history with potential catastrophic disruption, and spin rate/mutual orbit evolution via the YORP/Binary YORP (BYORP) effects, respectively. The MASs types are:

\begin{description}

\item[\emph{Type-1, T1: Large asteroids with small satellites}]
The T1 MASs have a large primary ($D_p > 90$ km) with one or several small satellites ($D_s < D_p/5$) in low-to-zero eccentricity ($e < 0.3$) orbits that are limited in system semi-major axis  $a/R_p < 15$, where $a$ is the semi-major axis of the mutual orbit and $R_p$ is the radius of the primary.  An example of a T1 MAS is 243 Ida with its satellite Dactyl.  The proposed formation scenarios for T1 systems are satellites generated either as ejecta from a large impact \citep{Durda:2004} or as material removed from the primary in a rotational mass shedding even \citep{Descamps:2008}.  These systems are expected to have anywhere from near zero to high internal porosity.

\item[\emph{Type-2, T2: Similar size double asteroids}] 
The T2 MASs are double asteroid systems comprised of two near equally-sized components with $a/R_{eq}$ $\approx 3-8$, where $R_{eq}$ is the radius of a sphere with the equivalent volume as the MAS.  The components have nearly circular orbits around their center of mass are tidally synchronized.  An example of a T2 MAS is 90 Antiope.  The T2 MASs are hypothesized to be the product of rotational fission as all of the observed systems cluster along the rotational fission equilibrium sequence \citep[for details see ][]{Descamps:2008}.  Rotational fission requires that the progenitor body has a low internal strength with little cohesive strength.  As such, the T2 MASs are expected to have a high internal porosity.

\item[\emph{Type-3, T3: Small, asynchronous systems}]
The T3 MASs are small, asynchronous systems with $D_{p,s} \lesssim 10$ km with small mass ratios. These systems can have a high eccentricity secondary, and they can also potentially be ternary (two satellite) systems.  An example of a T3 MAS is 1509 Esclangona.  The T3 MASs are hypothesized to form as various parts of an evolutionary sequence driven by 'rubble-pile' physics described for near earth asteroids (NEAs) \citep{Jacobson:2011rot}.  As part of the `rubble-pile' evolutionary model, these systems are expected to be collisionally evolved and have high internal porosity.

\item[\emph{Type-4, T4: Contact-binary asteroids}] 
The T4 MASs are contact binaries with two distinguishable components in a dumbbell-like shape.  An example of a T4 MAS is the Jupiter Trojan asteroid 624 Hektor.  These systems are hypothesized to form as either part of the same rotational fission sequence that forms T2 systems, or as double systems whose mutual orbits have decayed to the point of contact via the binary Yarkovsky-O'Keefe-Radziefvskii-Paddack (BYORP) effect \citep{Descamps:2008,Jacobson:2011rot}.  Both of these scenarios require low internal strength, so high internal porosity is expected for T4 MASs.
\end{description}

%The formation scenarios for MAS can be divided into three broad mechanisms:  1) those driven by impact processes; 2) those driven by `rubble-pile' physics and rotational effects, e.g., rotational fission and mass-shedding; and 3) those driven by tidal disruption.  

The formation scenarios for MASs can be divided into two broad mechanisms:  1) those driven by impact processes; and 2) those driven by `rubble-pile' physics and rotational effects, e.g., rotational fission and mass-shedding.  These mechanisms can operate individually or be interconnected.  For example, an impact can eject material that can remain in orbit forming a T1 MAS (small impactor on large target) or T3 MAS (nearly equal-sized, small impactor and target).   Such an impact can fracture the asteroid imparting a moderate macro-porosity of up to 30\% \citep{Britt:2002} to the primary body.  If sufficiently large, an impact can cause the catastrophic disruption of an asteroid, which can re-aggregate as a rubble-pile with large macro-porosity ($> 30\%$) \citep{Michel:2001,Britt:2002,Richardson:2005,Tanga:2009,Jacobson:2011rot}.  The rubble-pile asteroid, through rotational spin-up via the YORP effect, can then evolve into a T2 MAS through rotational fission and may further evolve into a T4 MAS through BYORP driving orbital decay.  Alternatively, it can evolve into a T1 or T3 MAS through mass-shedding.

The discerning factor between the formation scenarios is the internal structure, or rather the amount of vacuum space within the body, i.e., internal porosity.  The determination of porosity requires a measurement of the MAS density, and the identification of an appropriate meteorite analog with a known density.  In turn, knowledge of the meteorite analog requires, at a minimum, an accurate determination of taxonomic classification \citep[e.g.,][]{DeMeo:2009}, but it often requires a visible and NIR spectral analysis to determine mineralogical information, which is then related to specific meteorite analogs.
%an accurate taxonomic classification, and in the case of S-complex asteroids, requires accurate determination of the S-subtypes, S(I) - S(VII), defined by \citet{Gaffey:1993}.    

%The purpose of this study is to determine the taxonomy and mineralogy for a large sample of MAS to provide the required context to investigate the potential formation mechanisms.  What is required for this is visible and NIR spectral data. 
%[Introduce the mineralogy through band parameter analysis here.  Important parts to introduce:  NIR reflectance spectra contain diagnostic absorption features caused by Fe2+ crystal field transitions in ferrous iron-bearing silicates.  The 1~\micron \ feature is due to olivine and pyroxene while the 2~\micron \ feature is due to pyroxene only implying that ratios between the band areas are sensitive to the olivine-to-pyroxene ratio.  Band I center is controlled by iron-content and olivine-to-pyroxene content.  As such, the location of the Band I center can be used to determine the molar percentage of ferrosilite (Fs) in pyroxene and fayalite (Fa) in olivine.]

%\subsection{Mineralogy from spectral data}
Visible and NIR (0.45~--~2.5~\micron) reflectance spectra of S-complex asteroid surfaces contain a suite of diagnostic absorption features due to the mafic minerals, olivine and pyroxene.  These features are sensitive to relative and modal abundances as well as physical characteristics such as grain size, and thus carry compositional and mineralogical information.  Using two-component mixtures of olivine and pyroxene, \citet{Cloutis:1986} performed a landmark study that connected the band parameters (center, area, depth, and ratio of band areas) of the 1- and 2-\micron \ absorption bands (Band I and II, respectively) observed in iron-bearing silicates to the relative abundances of olivine and pyroxene and the modal abundances for the olivine and pyroxene phases.  Specifically, \citet{Cloutis:1986} demonstrated that: 1) the band area ratio (BAR),  Band II Area divided by Band I Area, is a grain-size independent, sensitive indicator of the olivine-to-orthopyroxene abundance; and 2) a comparison between Band I center location to the BAR is sensitive to the modal abundances of olivine and pyroxene.

%Band I being sensitive to olivine and pyroxene while Band II being only sensitive to pyroxene indicates that the BAR parameter is sensitive to the relative abundances of olivine and pyroxene.  As such, the BAR parameter is effective for determining the olivine ratio (ol. ratio = olivine / [olivine + pyroxene]) from spectral data for asteroids containing ferrous iron.
These band parameter sensitivities can be understood by examining the differences in the crystal field absorptions of olivine and pyroxene.  Olivine has a broad absorption feature near 1~\micron \ that is the composite of three distinct absorption bands.  This composite feature is due to \Ftwo \ electronic transitions occupying the M1 and M2 crystallographic sites of olivine \citep{Burns:1970}. Pyroxene has two absorption bands located at approximately 1 and 2~\micron.  The absorptions are caused by crystal field transitions of \Ftwo \ that typically occupy the M2 crystallographic site in pyroxene \citep{Clark:1957,Burns:1970}.  Hence, Band I is due to a combination of olivine and pyroxene while Band II is only due to pyroxene.  This differential absorption behavior indicates that a parameterization comparing Band I to Band II, in this case the BAR, is sensitive to the relative abundance of olivine to pyroxene, or the ratio ol/(ol~+~px) [hereafter ol ratio].

The wavelength position of the minimum of the absorption band, referred to as the band center, is effective for determining the modal abundances (mol\% Fe, Mg, Ca) because the position of the bands are dependent upon cation content.  For both olivine and low-calcium (Ca) pyroxene (Wollastonite (Wo), CaSiO$_3$,  $\lesssim 11$) there is a well-defined relationship between the Band I center position and ferrous iron content, such that the Band I and II centers are measured at longer wavelengths for increasing iron-content.  The relationship between band center and modal abundance is more complicated for high-Ca pyroxenes since increases in both \Ftwo \ and \Ctwo \ increase the wavelength position of both Band I and II \citep{Adams:1974,Burns:1972,Cloutis:1986}.  However, for single species pyroxenes, with low- or high-Ca content, the mol\% Wo versus mol\% Ferrosilite (Fs, Fe$_2$Si$_2$O$_6$) can be disentangled using the center position of Band II \citep{Gaffey:2002,Gaffey:2007,Burbine:2007,Burbine:2009}.  This relationship does not work for multi-assemblage pyroxenes, such as Ordinary Chondrite (OC) meteorites, because the central band position is a weighted average of the band positions of each pyroxene phase \citep{Gaffey:2007}.

In \S\ref{sec:obs}, we provide the technical details of the instruments and telescopes used in our MAS spectral survey along with a description of the data processing that was applied to the visible and NIR spectral data.  In \S\ref{sec:bdm} we introduce the Bus-DeMeo taxonomy and classify the MASs in our survey.  In \S\ref{sec:mineralogy} we describe the band parameter analysis routine and apply it to determine the \citet{Gaffey:1993} S-subtypes; mineralogy of S-complex and V-type asteroids; and identify meteorite analogs through the application of calibration studies.  The mineralogy and meteorite analog results are then applied in \S\ref{sec:macroporosity} to constrain the internal porosity of the MASs. % A statistical evaluation to find correlations between MAS type, mineralogy, and internal structure is presented in \S\ref{sec:mas_stats}.  
In \S\ref{sec:disc}, we discuss the results with respect to the MAS formation and evolution hypotheses.  We also explore a connection between LL chondrite-like mineralogies and Flora family membership.

%\subsection{Asteroid classifications}
%\label{sec:asteroid_classes}

%%%%%%%%%%%%%%%%%%%%%%%%%%%%%%%%%%%%%%%%%%%%%%%%%%%%
\section{MAS Data acquisition and processing}
\label{sec:obs}

Visible and/or near-IR (NIR) reflectance spectra of 42 MASs covering 0.45~--~2.5~\micron \ were obtained as part of an observational campaign started in August 2008, using two telescope facilities: The 4.1 m Southern Astrophysical Research (SOAR) Telescope using the Goodman High Throughput Spectrometer \citep{Clemens:2004} and NASA's 3 m Infrared Telescope Facility (IRTF) using the SpeX spectrograph \citep{Rayner:2003spex}. The observation details for the visible and near-infrared observations are given in Table~\ref{tab:mas_obs_soar} and \ref{tab:mas_obs_irtf}, respectively.  These tables provide the date of observation, observation start time (UT), number of observations $\times$ exposure time, heliocentric distance ($r_h$) at the time of observation, phase angle ($\alpha$), visible magnitude ($m_{\rm v}$), and  airmass of MAS. Tables~\ref{tab:mas_obs_soar} and \ref{tab:mas_obs_irtf} also provide each target MAS's standard star observation that was used for reduction and the B - V and V - K color of the standard star.

 %A summary of the MAS observations is provided in Table~\ref{tab:mas_obs_all}, and it includes the asteroid ID \#, asteroid name, Bus-DeMeo (BDM) Taxonomy, binary type, \citet{Demeo:2013} dynamical zone, and data reference.  

To obtain reflectance spectra and minimize the effects of atmospheric telluric absorption, standard stars are observed either immediately before or after the target observations.  The standards were selected to have similar airmass, spectral type, B - V color, and V - K color to the Sun.  The B, V, and K magnitudes and stellar spectral type were obtained from the SIMBAD database\footnote{http://simbad.u-strasbg.fr.simbad/}(and references therein).  The final relative reflectances for the MAS sample are calculated by dividing the spectrum of the object by the spectrum of the standard star for that observation.
%In order to maximize the quality of the spectral data, the MAS observations were made as close to meridian as possible to attain the lowest possible airmass.  The observations of standard stars were made either immediately before or after object observation, and they were selected to best match the airmass of the asteroid.  The standard stars are all of similar spectral type, B - V color, and V - K color to the Sun (see Tables~\ref{tab:mas_obs_soar} and \ref{tab:mas_obs_irtf}).

%%%%%%%%%%%%%%%%%%%%%%%%%%%%%%%%%%%%%%%%%%%%%%%%%%%%
\subsection{Visible spectra}
\label{sec:obs_vis}

The visible SOAR spectral campaign includes 25 MASs covering 0.45~--~0.85~\micron \ obtained over 15 nights spanning December 2011 - October 2012.  All observations were made using a 1.35" wide slit with the 300 l/mm grating giving a maximum resolving power, R=1390.  The observational details for the SOAR-Goodman are listed in Table~\ref{tab:mas_obs_soar}.

The SOAR + Goodman spectral data reduction is performed using standard techniques with minor modifications due to the unique nature of the Goodman instrument.  The one-dimensional spectra for the MAS and standard star frames are extracted from the two dimensional frames by summing pixels in the data range and subtracting off the background value for each column.  The background for each column is measured as the mean of non-target pixels for each column.   For each of the extracted spectra, we reject cosmic ray and other anomalous signal detections by removing pixels that are 5-$\sigma$ outliers within a 100 pixel wide box that steps through the spectrum.  The bad pixels are replaced with a linear interpolation of the removed data range.  This bad pixel rejection process is iterated until no 5-$\sigma$ detections are found.  The cleaned spectra for the standard stars and MASs are then shifted to a common reference\footnote{For SOAR + Goodman, spectral shift can be up to several pixels.}, and co-added to produce a composite star and MAS frame.  The relative reflectance spectrum for a MAS is then calculated by dividing the co-added object frame by the co-added standard star frame.  We select standard stars that have colors similar to the Sun, and are temporally and spatially observed near the MAS observations, to minimize the effects of the telluric absorptions due to Earth's atmosphere. 

Flat-fielding is left out of the procedure because of the known scattered light contamination off the back of the second filter wheel (http://ctio.naoa.edu).  This scattered light introduces a systematic noise pattern that cannot be removed from the flat field images, precluding them from the reduction pipeline.  However, by placing the star and object at the same position on the CCD, the flat fielding is intrinsically incorporated into our reduction via the division of the object by the star.   The wavelength calibration is determined using mercury and argon lamp spectra.  The finalized visible spectra are determined by binning the data with a 5-pixel box and normalizing to the reflectance value at 0.55~\micron.  The observational uncertainties are taken to be the standard deviation of the reflectance values within each bin.

%The general reduction procedure is to extract the 1D spectrum for all object and star frames, remove bad pixel and cosmic ray hits from the 1D spectra, shift\footnote{For SOAR + Goodman, spectral shift can be up to several pixels.} the object and star spectra to a common reference, co-add the star and object frame, and finally divide the composite object spectrum by the composite star spectrum to obtained a relative reflectance spectrum for the MAS.  The use of a standard star that is as close to a solar spectrum as possible is required to remove telluric absorptions from Earth's atmosphere.  Flat-fielding is left out of the procedure because of the known scattered light contamination off the back of the second filter wheel (http://ctio.naoa.edu).  The scattered light introduces a systematic noise pattern that cannot be removed from the flat field images.  However, by placing the star and object at the same position on the CCD, the flat fielding is intrinsically incorporated into our reduction via the division of the object by the star.   The wavelength calibration is applied using Mercury and Argon lamp spectra.  The finalized visible spectra are determined by binning with a 5-pixel box and normalizing to the reflectance value at 0.55~\micron.  The observational uncertainties are taken to be the standard deviation of the reflectance values within each bin.

Additional processing to the SOAR + Goodman spectral data is required because it contains significant fringing effects longward of 0.65~\micron.  We attempt to correct the fringing effects through the application of a Fast Fourier Transform (FFT) filter that isolates the fringing frequencies.  The FFT filter is applied to the one-dimensional spectra, prior to wavelength calibration, of the individual star and object frames to avoid convoluting the slight differences in fringing patterns between individual spectral frames.  We tune the FFT filter to retain high frequencies (less than 6 pixels in wavelength), which are due to observational uncertainties; and to retain the lower frequencies (greater than 18 pixels in wavelength), which are representative of the true spectral features.   %An example of a Soar + Goodman object frame before and after filtering is shown in Fig.~\ref{fig:soar_fringing}.  
We further mitigate the effects of fringing by extending our IRTF spectra down to 0.75~\micron \ to replace the 0.75~--~0.85~\micron \ portion of SOAR + Goodman data.

%\begin{figure}[!h]
%\begin{center}
%\includegraphics[width=8cm]{fig02.eps}
%\caption{The raw 1D spectrum of 762 Pulcova extracted from the SOAR + Goodman High Throughput Spectrometer data both with (\emph{black}) and without (\emph{red}) the fringe correcting Fast Fourier Transform (FFT) filter applied. The fringing pattern is apparent longwards of 0.75~\micron \ is apparent in the unfiltered spectrum.}
%\label{fig:soar_fringing}
%\end{center}
%\end{figure}

\begin{sidewaystable} \scriptsize \renewcommand{\arraystretch}{0.8} \vspace{0.005in}
\caption{SOAR Goodman Observations}
\begin{center}
\begin{tabular}{ p{0.65cm} p{1.7cm} p{0.9cm} p{1.5cm} p{1.1cm} p{0.7cm} p{0.7cm} p{0.7cm} p{1.6cm}  p{0.9cm} p{0.9cm} p{0.9cm} } 

\toprule
ID  & Date & Time & texp(s) & r$_h$(AU) & $\alpha^o$ & $m_{\rm v}$ & Airmass & Standard & B - V & V - K & Spectral \\
 &  & (UT) &  &  &  &  & [Object] & Star & [Star] & [Star] & Type \\ \toprule
121 & 8-Jul-12 & 19:48.0 & 3 x 100 & 3.90 & 15.0 & 14.0 & 1.38 & HD 112280 & 0.69 & 1.57 & G5 \\
130 & 14-Jun-12 & 05:51.6 & 3 x 100 & 3.67 & 12.9 & 13.1 & 1.39 & HD 135264 & 0.54 & -- & G0V \\
283 & 8-Jul-12 & 36:33.2 & 3 x 150 & 3.45 & 16.8 & 15.0 & 1.21 & HD 104576 & 0.70 & 1.82 & G3V \\
702 & 27-Jul-12 & 49:58.1 & 3 x 120 & 3.13 & 14.3 & 12.4 & 1.01 & HD 155827 & 0.57 & 1.41 & G1V \\
762 & 7-Jul-12 & 19:11.8 & 3 x 100 & 3.18 & 17.3 & 13.8 & 1.23 & HD 122998 & 0.60 & 1.60 & G1V \\
939 & 7-Oct-12 & 52:13.4 & 3 x 40 & 2.58 & 17.5 & 17.6 & 1.45 & BD+13 213 & 0.89 & 1.37 & G5 \\
1089 & 12-Jun-12 & 08:20.2 & 3 x 140 & 2.41 & 16.4 & 15.5 & 1.08 & HD 189764 & 0.70 & 1.43 & G1V \\
1333 & 27-Jul-12 & 31:21.3 & 3 x 210 & 2.99 & 14.5 & 16.3 & 1.08 & HD 157842 & 0.80 & 1.91 & G2V \\
1717 & 7-Oct-12 & 09:39.0 & 3 x 125 & 2.00 & 14.8 & 16.8 & 1.31 & HD 2680 & 0.66 & 1.52 & G5 \\
2577 & 27-Jul-12 & 08:29.3 & 3 x 250 & 2.11 & 28.6 & 17.3 & 1.47 & HD 138952 & 0.44 & 1.24 & G0 \\
3309 & 12-Jun-12 & 47:37.7 & 3 x 215 & 1.73 & 28.3 & 15.9 & 1.17 & HD 201454 & 0.66 & 1.36 & G2V \\
3703 & 28-Jan-12 & 41:57.2 & 2 x 2340 & 2.60 & 7.1 & 17.7 & 1.46 & BD+12 1603 & 0.60 & 1.59 & G5 \\
3782 & 9-Jul-12 & 24:18.6 & 3 x 350 & 2.32 & 19.2 & 16.2 & 1.22 & HD 208598 & 0.54 & -- & G0 \\
4029 & 10-Sep-11 & 59:57.6 & 3 x 480 & 2.69 & 5.7 & 16.6 & 1.24 & G 31-2 & 0.64 & 1.49 & G2V \\
4492 & 10-Sep-11 & 55:38.4 & 3 x 210 & 2.38 & 7.7 & 16.0 & 1.35 & BD+03 2 & 0.59 & 1.53 & G0 \\
4674 & 7-Jul-12 & 01:39.8 & 3 x 125 & 1.97 & 23.7 & 16.6 & 1.21 & HD 144821 & 0.60 & 1.41 & G2V \\
4786 & 10-Dec-11 & 44:21.2 & 3 x 1075 & 2.81 & 12.0 & 17.6 & 1.27 & BD+09 405 & 0.25 & 1.83 & G0 \\
5481 & 7-Jul-12 & 38:22.6 & 3 x 250 & 2.22 & 20.4 & 16.7 & 1.21 & HD 139732 & 0.56 & 1.32 & G2V \\
6244 & 27-Jul-12 & 45:51.3 & 3 x 280 & 2.11 & 10.6 & 16.1 & 1.01 & HD 111017 & 0.66 & 1.42 & G1V \\
6708 & 19-Jan-12 & 25:20.7 & 3 x 670 & 2.89 & 5.5 & 17.2 & 1.61 & LTT 12322 & 0.61 & 1.45 & G5 \\
7225 & 7-Jul-12 & 42:13.4 & 3 x 210 & 2.38 & 25.1 & 17.6 & 1.38 & HD 180702 & 0.54 & 1.42 & G1V \\
9617 & 9-Sep-11 & 03:28.6 & 3 x 430 & 2.18 & 12.1 & 17.5 & 1.21 & HD 2818 & 0.56 & 1.61 & G5 \\
17246 & 29-Aug-11 & 55:55.6 & 1 x 1585 & 2.85 & 10.1 & 18.1 & 1.17 & HD 224448 & 0.60 & 1.50 & G0 \\
17260 & 7-Oct-11 & 04:57.4 & 3 x 535 & 2.27 & 5.5 & 17.0 & 1.31 & BD+08 205 & 0.65 & 1.47 & G5 \\
32008 & 18-Jan-12 & 40:03.0 & 3 x 1400 & 2.35 & 8.7 & 17.6 & 1.25 & HD 43872 & 0.66 & 1.54 & G5 \\ \bottomrule  

%\mbox{footnotes}

\end{tabular}
\label{tab:mas_obs_soar}
\end{center}
\end{sidewaystable}

%%%%%%%%%%%%%%%%%%%%%%%%%%%%%%%%%%%%%%%%%%%%%%%%%%%%
\subsection{Near-infrared spectra}
\label{sec:obs_nir}

The NIR IRTF SpeX spectral campaign includes 34 MASs covering 0.7 - 2.5~\micron \ obtain over 12 nights spanning March 2007 - May 2013.  The NIR spectra were obtained using SpeX in the low-resolution ($R = 250$) prism mode with a 0.8" slit width.  The observational details for the IRTF-SpeX observations are listed in Table~\ref{tab:mas_obs_irtf}.

The spectra were acquired in a standard ABBA nodding sequence.  Data frames were taken in pairs with the object dithered along the slit between an A and B position.   Generally, two ABBA sequences are taken on target followed by two ABBA sequences on a standard solar-type (G-dwarf) star.   To minimize atmospheric extinction effects, the standard stars are chosen to be within four degrees of the target.  The IRTF flat field images are obtained by illuminating an integrating sphere that is in the calibration box attached to the spectrograph.  This box also contains an argon lamp that was used to perform wavelength calibrations.

The NIR reflectance data were processed using standard techniques following \citet{Emery:2003}.  Bad pixel masks were created using dark and flat-field frames to identify pixels whose response is consistently low, high, or erratic between frames.  Image pairs are subtracted to produce a first-order removal of the sky background emission.  The background subtracted images are then divided by a flat-field image to remove pixel-to-pixel variation across the chip.  The frames are also searched for cosmic ray hits and other anomalous signal detections.  One-dimensional spectra are then extracted from the two-dimensional frames by summing pixels in the data range for each column on the chip.

The one-dimensional spectra are still contaminated by telluric absorptions.  These are corrected for by dividing the spectrum of an extracted MAS spectrum by a standard star with the same airmass (+/- 0.05).  This division also corrects the asteroid spectra for the shape of the solar spectrum.  Typically, multiple ABBA sequences for each asteroid are taken and summed together in order to increase the signal-to-noise (S/N) ratio.  The error bars are the standard deviation of the mean of the co-added frames.

\begin{sidewaystable} \scriptsize \renewcommand{\arraystretch}{0.8} \vspace{0.005in}
\caption{IRTF SpeX Observations}
\begin{center}
\begin{tabular}{ p{0.6cm} p{1.7cm} p{0.9cm} p{1.5cm} p{1.1cm} p{0.7cm} p{0.7cm} p{0.7cm} p{1.7cm} p{0.9cm} p{0.9cm} p{0.9cm} } 

\toprule
ID  & Date & Time & t$_{\rm exp}$(s) & r$_h$(AU) & $\alpha^o$ & $m_{\rm v}$ & Airmass & Standard & B - V & V - K & Spectral \\
 &  & (UT) &  &  &  &  & [Object] & Star & [Star] & [Star] & Type \\ \toprule
45 & 13-Sep-07 & 6:30:00  & 14x15 & 2.61 & 5.4 & 11.06 & 1.28  & HD 197400 & 0.60 & 1.48 & G3V  \\
107 & 29-Mar-07 & 6:05:00 & 8x40 & 3.22 & 16.1 & 12.70 & 1.01 & HD 69809 & 0.67 & 1.49 & G0 \\
121 & 1-Apr-11 & 5:40:00 & 16x120 & 3.34 & 16.1 & 13.60 & 1.06 & SAO 97748 & 0.66 & 1.52 & G0 \\
130 & 14-Aug-08 & 10:48:00 & 14x15 & 2.60 & 9.2 & 10.90 & 1.25 & HD 219432 & 0.74 & 1.42 & G0 \\
216 & 14-Aug-08 & 11:22:00 & 14x15 & 2.29 & 16.4 & 10.60 & 1.03 & HD 220848 & 0.67 & 1.59 & G5 \\
283 & 13-Nov-10 & 15:30:00 & 12x120 & 3.19 & 17.9 & 14.50 & 1.01 & HD 247554 & 0.63 & 1.46 & G5 \\
379 & 30-Mar-07 & 14:22:00 & 12x120 & 3.50 & 9.9 & 14.40 & 1.17 & HD 95622 & 0.63 & 1.33 & G5 \\
762 & 14-Aug-08 & 10:09:00 & 14x25 & 3.46 & 3.7 & 13.30 & 1.11 & HD 210392 & 0.71 & 1.44 & G0 \\
809 & 29-Mar-07 & 6:56:00 & 12x120s & 2.72 & 8.8 & 16.10 & 1.17 & HD 95622  & 0.63 & 1.33 &  G5 \\
809 & 14-Aug-08 & 8:26:00 & 26x90 & 1.95 & 19.1 & 14.70 & 1.26 & HD 175276 & 0.55 & 1.72 & G2,3V\\
939 & 15-Apr-13 & 8:30:00 & 30x120 & 2.63 & 19.0 & 16.70 & 1.18 & SAO 98763 & 0.64 & 1.49 & G0 \\
1089 & 1-Apr-11 & 9:15:00 & 24x120 & 1.94 & 14.4 & 16.10 & 1.03 & HD 101588 & 0.59 & 1.52 & G5 \\
1313 & 6-Jul-11 & 13:00:00 & 32x120 & 2.36 & 23.9 & 15.90 & 1.35 & SAO 128027 & 0.66 & 1.48 & G0 \\
1333 & 10-Apr-12 & 14:30:00 & 8x120 & 2.96 & 18.8 & 16.70 & 1.20 & HD 165479 & 0.62 & 1.69 & G0 \\
1509 & 6-Jul-11 & 7:00:00 & 24x120 & 1.85 & 15.9 & 14.60 & 1.81 & HD 150952 & 0.65 & 1.49 & G1V \\
1717 & 9-May-13 & 8:50:00 & 16x120 & 2.27 & 4.7 & 14.90 & 1.27 & HD 120882 & 0.62 & 1.42 & G2V \\
2131 & 15-Apr-13 & 7:00:00 & 28x120 & 2.10 & 27.0 & 16.50 & 1.31 & HD 75705 & 0.62 & 1.48 & G5V \\
2577 & 10-Apr-12 & 11:40:00 & 32x120 & 1.97 & 18.8 & 15.80 & 1.06 & HD 165479 & 0.62 & 1.69 & G1 \\
3309 & 13-Nov-10 & 11:45:00 & 16x120 & 1.91 & 8.1 & 15.40 & 1.11 & HD 247554 & 0.63 & 1.46 & G5 \\
3623 & 13-Nov-10 & 13:20:00 & 16x120 & 2.71 & 1.7 & 15.60 & 1.45 & SAO 93559 & 0.70 & 1.48 & G0 \\
3749 & 13-Sep-10 & 6:30:00 & 32x120 & 2.43 & 15.7 & 16.90 & 1.30 & HD 197400 & 0.60 & 1.48 & G3V \\
3782 & 1-Jul-12 & 14:00:00 & 4x120 & 2.32 & 21.4 & 16.40 & 1.01 & HD 191320 & 0.61 & 1.48 & G0 \\
3982 & 14-Aug-08 & 6:50:00 & 14x120 & 1.78 & 31.2 & 15.60 & 1.32 & HD 152931 & 0.66 & 1.54 & G3V \\
4607 & 15-Apr-13 & 14:40:00 & 32x120 & 2.29 & 19.1 & 16.60 & 1.41 & HD 149826  & 0.66 & 1.48 & G3V \\
4674 & 13-Sep-10 & 9:20:00 & 16x120 & 1.92 & 17.3 & 16.00 & 1.02 & HD 215274 & 0.67 & 1.56 & G5V \\
4786 & 9-May-13 & 6:00:00 & 28x120 & 2.11 & 25.0 & 16.80 & 1.01 & BD+20 2449 & 0.65 & 1.52 & G0 \\
5407 & 13-Sep-10 & 10:20:00 & 16x120 & 2.04 & 9.5 & 16.20 & 1.29 & SAO 147139 & 0.62 & 1.23 & G1V \\
5481 & 1-Jul-12 & 6:50:00 & 2x120 & 2.22 & 18.5 & 16.60 & 1.50 & SAO 206536 & 0.61 & 1.48 & G0 \\
5905 & 6-Jul-11 & 12:50:00 & 32x120 & 1.78 & 33.4 & 17.10 & 1.43 & HD 224908 & 0.62 & 1.47 & G5 \\
6265 & 9-May-13 & 9:50:00 & 24x120 & 2.26 & 6.4 & 16.20 & 1.19 & HD 120882 & 0.62 & 1.42 & G2V \\
6708 & 13-Sep-10 & 13:40:00 & 24x120 & 2.23 & 21.4 & 16.70 & 1.10 & HD 19451 & 0.73 & -- & G5 \\
7225 & 11-Mar-12 & 12:00:00 & 8x120 & 2.11 & 17.2 & 16.00 & 1.35 & SAO 82672 & 0.61 & 1.57 & G0V \\
8116 & 9-May-13 & 11:30:00 & 30x120 & 2.51 & 13.8 & 17.50 & 1.35 & SAO 183790 & 0.61 & 1.47 & G2V \\
17260 & 9-May-13 & 13:00:00 & 12x120 & 2.20 & 17.9 & 17.50 & 1.34 & HD 142963 & 0.63 & 1.49 & G3V \\
34706 & 15-Apr-13 & 12:40:00 & 30x120 & 2.75 & 21.0 & 19.70 & 1.36 & HD 115273 & 0.63 & 1.48 & G0   \\ \bottomrule

%\mbox{footnotes}

\end{tabular}
\label{tab:mas_obs_irtf}
\end{center}
\end{sidewaystable}

%%%%%%%%%%%%%%%%%%%%%%%%%%%%%%%%%%%%%%%%%%%%%%%%%%%%
\subsection{Visible plus near-infrared spectra}
\label{sec:obs_visnir}

In order to accomplish a robust taxonomic determination for as many of our MASs as possible, we combine the visible and near-infrared spectra for MASs with both SOAR Goodman and IRTF SpeX data.  We supplement our MAS data set with visible spectra from the Small Main-Belt Asteroid Spectroscopic Survey, Phase II (SMASS II) database \citep{Bus:2002a} for 45 Eugenia, 107 Camilla, 216 Kleopatra, 809 Lundia, 2131 Mayall, 4607 Seilandfarm, and 5407 1992 AX;  and from the Small Solar System Objects Spectroscopic Survey (S3OS2) dataset \citep{Lazzaro:2007} for 1509 Esclangona.  We also supplement our data set with an infrared spectrum from \citet{Moskovitz:2010} for 3703 Volkonskaya.  In total, we have 26 visible+NIR reflectance spectra, 9 NIR only spectra, and 7 visible only spectra for a total sample of 42 MAS spectra.

The visible and NIR spectra are combined by fitting 3$^{rd}$ order polynomials to the red and blue ends of the visible and NIR spectra, respectively.  The NIR reflectance values are then scaled to match the normalized visible reflectances in a wavelength region within the 0.75~--~0.8~\micron \ range.  At 0.75~\micron, the NIR data is typically more reliable than the fringe-affected SOAR-Goodman data, and as such, the combined spectra is taken as visible data up to 0.75~\micron \ and IR data past 0.75~\micron.  The NIR reflectance errors are propagated by multiplying by the scaling factor.  Finally, all of the spectra with visible data are normalized to unity at 0.55~\micron.  If only NIR spectral data are available, the finalized spectra are normalized to unity at 1.0~\micron. The MAS visible, NIR, and visible + NIR reflectance spectra are presented in Fig.~\ref{fig:spectra_thumbs}.

%%%%%%%%%%%%%%%%%%%%%%%%%%%%%%%%%%%%%%%%%%%%%%%%%%%%
%%%%%%%%%%%%%%%%%%%%%%%%%%%%%%%%%%%%%%%%%%%%%%%%%%%%
\section{Taxonomy classification}  
\label{sec:bdm}

We classify our reflectance spectra that have NIR or visible plus NIR data using the Bus-DeMeo taxonomy \citep{DeMeo:2009}.  The Bus-DeMeo taxonomy extends the visible light-based Bus taxonomy \citep{Bus:1999,Bus:2002a,Bus:2002b} to include  NIR data  ($< 2.5$~\micron) of asteroids.  The extension into the NIR allows the spectra of asteroids to be separated, via a principle component analysis, into 24 different classes within four groups:  S-complex, C-complex, X-complex, and End Members \citep{DeMeo:2009}. Each of our MAS spectra are classified in the Bus-DeMeo taxonomy using MIT's Bus-DeMeo Taxonomy Classification Tool; the results are summarized in Table~\ref{tab:BDM}.  Out of the 42 MASs in our sample, we report 28 new Bus-DeMeo classifications.

In most cases, the online tool reports a single classification for each MAS spectrum.  In a few cases, and frequently when only NIR data are available, the classification will be non-unique, and the online tool will report multiple likely Bus-DeMeo classes.  For these cases, comparisons between the Bus-DeMeo classes' mean spectra\footnote{Mean Bus-DeMeo class spectra acquired from the SMASS MIT Bus-Demeo Taxnoomy website: http://smass.mit.edu/busdemeoclass.html} are made in an attempt to discern the taxonomy.  If the classification is ambiguous, all of the classes listed by the online classification tool are listed in Table~\ref{tab:BDM}.  For S-complex and V-type asteroid spectra, the sub-assignment of `w' is appended to the Bus-DeMeo taxonomic class if the spectral slope is above an arbitrarily determined cut-off value of 0.25 \citep{DeMeo:2009}.  The `w' is added to account for the catch-all process of space weathering, which has been shown to redden spectral slopes \citep{Clark:2002}. While \citet{DeMeo:2009} make no claim as to whether an assignment of `w' reflects weathering on the asteroid surfaces, the moniker potentially indicates an increased likelihood of a weathered surface.

\begin{table} \scriptsize \renewcommand{\arraystretch}{0.8} \vspace{0.005in}
\caption{Bus-Demeo Taxonomy of MAS Sample}
\begin{center}
\begin{tabular}{ p{0.5cm} p{1.9cm} p{1.8cm} p{1.5cm} p{0.7cm} p{1.3cm} p{1.0cm} p{1.8cm} } 

\toprule
\bf{ID}  & \bf{Name} & \bf{BDM}  & \bf{MAS Type} & \bf{a(AU)} & \bf{Population$^1$} & \bf{Family} & \bf{New BDM?} \\ \toprule
45 & Eugenia & C & T1-2moons & 2.72 & MMB & - & Yes \\
107 & Camilla & Xe & T1 & 3.50 & Cybele & - & Yes \\
121 & Hermione & Ch & T1 & 3.45 & Cybele & - & Yes \\
130 & Elektra & Ch & T1 & 3.12 & OMB & - & No \\
216 & Kleopatra & Xe & T1-2moons & 2.80 & MMB & - & No \\
283 & Emma & X & T1 & 3.05 & OMB & Eos & Yes \\
379 & Huenna & X-, C-complex & T1 & 3.14 & OMB & - & - \\
702 & Alauda & - & T1 & 3.19 & OMB & - & - \\
762 & Pulcova & C & T1 & 3.16 & OMB & - & Yes \\
809 & Lundia & V & T2 & 2.28 & IMB & Flora & Yes \\
939 & Isberga & Sw & U & 2.25 & IMB & Flora & Yes \\
1089 & Tama & Sw & T2 & 2.21 & IMB & Flora & Yes \\
1313 & Berna & Q Sq & T2 & 2.66 & MMB & - & Yes \\
1333 & Cevenola & Sr & T4 & 2.63 & MMB & Eunomia & Yes \\
1509 & Esclangona & Sw & T3 & 1.87 & Hungaria & Hungaria & Yes \\
1717 & Arlon & Sw & T3 & 2.20 & IMB & - & Yes \\
2131 & Mayall & Sw & T3 & 1.89 & Hungaria & Hungaria & Yes \\
2577 & Litva & Sw & T3 & 1.90 & Hungaria & Hungaria & Yes \\
3309 & Brorfelde & Sw & T3 & 1.82 & Hungaria & Hungaria & Yes \\
3623 & Chaplin & Q Sq & T4 & 2.85 & OMB & Koronis & Yes \\
3703 & Volkonskaya & V & T3 & 2.33 & IMB & Vesta & No \\
3749 & Balam & Q Sq & T3 & 2.24 & IMB & Flora & Yes \\
3782 & Celle & Vw & T3 & 2.41 & IMB & Vesta & Yes \\
3982 & Kastel & V & T3 & 2.26 & IMB & Flora & Yes \\
4029 & Bridges & - & T3 & 2.53 & MMB & - & - \\
4492 & Debussy & - & T2 & 2.77 & MMB & - & - \\
4607 & Seilandfarm & L & U & 2.26 & IMB & - & Yes \\
4674 & Pauling & Sw & T3 & 1.86 & Hungaria & Hungaria & Yes \\
4786 & Tatianina & Xe & T3 & 2.36 & IMB & - & Yes \\
5407 & 1992AX & S & T3 & 1.84 & Hungaria & - & No \\
5481 & Kiuchi & Vw & T3 & 2.34 & IMB & Vesta & Yes \\
5905 & Johnson & Q Sq & T3 & 1.91 & Hungaria & - & Yes \\
6244 & Okamoto & - & T3 & 2.16 & IMB & Flora & - \\
6265 & 1985TW3 & Q Sq L K & T3 & 2.17 & IMB & - & - \\
6708 & Bobbievaile & K & T3 & 2.44 & IMB & - & Yes \\
7225 & Huntress & S & T3 & 2.34 & IMB & - & Yes \\
8116 & Jeanperrin & Q Sq L K & T3 & 2.25 & IMB & Flora & - \\
9617 & Grahamchapmin & - & T3 & 2.22 & IMB & Flora & - \\
17246 & 2000GL74 & - & T2 & 2.84 & OMB & Koronis & - \\
17260 & 2000JQ58 & S & T3 & 2.20 & IMB & - & Yes \\
32008 & 2000HM53 & - & T3 & 2.19 & IMB &  & - \\
34706 & 2001OP83 & V & T3 & 2.25 & IMB & Vesta & Yes  
 \\ \bottomrule

\vspace{0.075cm}
\mbox{BDM is Bus-DeMeo Taxa ~~~~~U is Unclassified MAS Type}
\mbox{(1) Population refers to the zones defined in \citet{Demeo:2013}:}
\mbox{IMB, MMB, OMB - Inner, Middle, Outer Main Belt}

\end{tabular}
\label{tab:BDM}
\end{center}
\end{table}

%%%%%%%%%%%%%%%%%%%%%%%%%%%%%%%%%%%%%%%%%%%%%%%%%%%%
%%%%%%%%%%%%%%%%%%%%%%%%%%%%%%%%%%%%%%%%%%%%%%%%%%%%
\section{Mineralogical analysis}
\label{sec:mineralogy}

Using the custom IDL (Interactive Data Language) routine, the Spectral Analysis Routine for Asteroids or SARA\footnote{SARA is freely available via GitHub: https://github.com/SeanSLindsay/SARA.git}, developed by the authors, we perform a spectral band analysis on the S- and V-type (n = 24) asteroids in our sample to determine the Band I and II centers, areas, and depths as well as the Band I slope.   The methodology of the algorithm is described in \S\ref{subsec:ba_code}.  The measured band parameters are applied to: 1) parse the S-type asteroids into their subtypes, as defined by \citet{Gaffey:1993}; and 2) to derive mineralogical information and identify meteorite analogs for the S-complex and V-type MAS asteroids (\S\ref{subsec:ba_ssub}).  The synthesis of these results is applied in \S\ref{sec:oc_min} and \S\ref{sec:vtype_min} in an attempt to determine the meteorite subtypes of the OC and HED meteorite groups.

%%%%%%%%%%%%%%%%%%%%%%%%%%%%%%%%%%%%%%%%%%%%%%%%%%%%
\subsection{Band Analysis Methods} 
\label{subsec:ba_code}

In this section, we describe the methodology of the IDL-based band parameter analysis routine called the Spectral Analysis Routine for Asteroids (SARA).  The IDL code is designed to be applicable to any near-IR (plus visible, if available) reflectance spectral data set that contains Bands I and II due to the cation \Ftwo \ in the crystalline structure.  The methodology attempts to provide an available to the public, generalized band analysis routine.  Here we describe the SARA algorithm methodology and apply it to calculate the band parameters for our MAS sample.  The results of our band analysis are summarized in Table~\ref{tab:band_analysis}.

\begin{sidewaystable} \scriptsize \renewcommand{\arraystretch}{0.8} \vspace{0.005in}
\caption{Band Parameter Analysis}
\begin{center}
\begin{tabular}{ p{0.5cm} p{1.6cm} p{0.6cm} p{0.8cm} p{0.8cm} p{0.8cm} p{1.0cm} p{1.0cm} p{0.8cm} p{0.8cm} p{0.8cm} p{1.1cm} p{1.1cm} p{0.8cm} p{0.8cm} p{0.8cm} } 

\toprule
ID & Object & BDM & Temp & BIC & BIC & BI Area & BI Area & $\Delta$BIIC & BIIC & BIIC & BII Area & BII Area & $\Delta$BAR & BAR & BAR \\
 &  &  & (K) &  & Error &  & Error &  &  & Error &  & Error &  &  & Error \\ \toprule
809 & Lundia & V & 152.6 & 0.931 & 0.001 & 0.117 & 0.000 & 0.027 & 1.964 & 0.003 & 0.224 & 0.001 & -0.105 & 1.815 & 0.009 \\
939 & Isberga & S & 166.4 & 1.001 & 0.004 & 0.064 & 0.001 & 0.027 & 2.054 & 0.082 & 0.013 & 0.002 & -0.105 & 0.097 & 0.027 \\
1089 & Tama & S & 191.7 & 0.980 & 0.002 & 0.040 & 0.000 & 0.022 & 1.951 & 0.002 & 0.025 & 0.000 & -0.086 & 0.536 & 0.006 \\
1313 & Berna & Sq Q & 177.0 & 0.990 & 0.002 & 0.051 & 0.000 & 0.025 & 1.947 & 0.003 & 0.023 & 0.000 & -0.097 & 0.363 & 0.007 \\
1333 & Cevenola & S & 158.9 & 1.001 & 0.019 & 0.068 & 0.002 & 0.028 & 1.924 & 0.031 & 0.042 & 0.004 & -0.111 & 0.514 & 0.054 \\
1509 & Esclangona & S & 196.7 & 0.953 & 0.004 & 0.028 & 0.000 & 0.021 & 1.926 & 0.004 & 0.020 & 0.000 & -0.083 & 0.620 & 0.019 \\
1717 & Arlon & S & 176.6 & 0.983 & 0.004 & 0.062 & 0.001 & 0.025 & 1.937 & 0.005 & 0.017 & 0.000 & -0.098 & 0.173 & 0.007 \\
2131 & Mayall & S & 184.5 & 0.906 & 0.003 & 0.034 & 0.001 & 0.023 & 1.946 & 0.026 & 0.026 & 0.002 & -0.092 & 0.667 & 0.065 \\
2577 & Litva & S & 189.2 & 1.044 & 0.001 & 0.066 & 0.000 & 0.022 & 2.065 & 0.005 & 0.040 & 0.000 & -0.088 & 0.518 & 0.008 \\
3309 & Brorfelde & S & 192.2 & 0.955 & 0.003 & 0.040 & 0.000 & 0.022 & 1.912 & 0.005 & 0.038 & 0.001 & -0.086 & 0.854 & 0.025 \\
3623 & Chaplin & Sq Q & 162.6 & 0.918 & 0.005 & 0.030 & 0.001 & 0.027 & 2.004 & 0.059 & 0.028 & 0.002 & -0.108 & 0.814 & 0.071 \\
3703 & Volkonskaya & V & 171.3 & 0.927 & 0.002 & 0.080 & 0.001 & 0.026 & 1.959 & 0.008 & 0.153 & 0.002 & -0.102 & 1.817 & 0.034 \\
3749 & Balam & Sq Q & 176.3 & 0.982 & 0.004 & 0.076 & 0.001 & 0.025 & 1.983 & 0.031 & 0.034 & 0.002 & -0.098 & 0.353 & 0.027 \\
3782 & Celle & V & 157.8 & 0.936 & 0.002 & 0.090 & 0.000 & 0.028 & 1.950 & 0.005 & 0.161 & 0.002 & -0.112 & 1.677 & 0.020 \\
3982 & Kastel & V & 202.9 & 0.932 & 0.001 & 0.123 & 0.000 & 0.019 & 1.960 & 0.004 & 0.265 & 0.001 & -0.078 & 2.065 & 0.013 \\
4674 & Pauling & S & 182.6 & 1.007 & 0.003 & 0.034 & 0.000 & 0.023 & 1.946 & 0.015 & 0.015 & 0.001 & -0.093 & 0.339 & 0.022 \\
5407 & 1992ax & S & 190.4 & 0.920 & 0.004 & 0.037 & 0.001 & 0.022 & 1.907 & 0.079 & 0.026 & 0.002 & -0.087 & 0.965 & 0.058 \\
5481 & Kiuchi & V & 172.2 & 0.927 & 0.006 & 0.125 & 0.004 & 0.026 & 1.957 & 0.026 & 0.203 & 0.005 & -0.101 & 1.523 & 0.063 \\
5905 & Johnson & Sq Q & 194.4 & 0.964 & 0.004 & 0.040 & 0.000 & 0.021 & 1.947 & 0.005 & 0.027 & 0.001 & -0.084 & 0.601 & 0.022 \\
6265 & 1985tw3 & Sq Q & 174.9 & 0.993 & 0.004 & 0.061 & 0.000 & 0.025 & 1.944 & 0.013 & 0.022 & 0.001 & -0.099 & 0.258 & 0.015 \\
6708 & Bobbievaile & K & 186.3 & 1.059 & 0.005 & 0.037 & 0.000 & 0.023 & 2.114 & 0.220 & 0.001 & 0.001 & -0.090 & -0.059 & 0.032 \\
7225 & Huntress & S & 188.1 & 1.063 & 0.002 & 0.106 & 0.001 & 0.022 & 1.949 & 0.011 & 0.002 & 0.001 & -0.089 & -0.070 & 0.014 \\
8116 & Jeanperrin & Sq Q & 171.6 & 0.992 & 0.006 & 0.074 & 0.001 & 0.026 & 1.948 & 0.103 & 0.020 & 0.002 & -0.101 & 0.164 & 0.027 \\
17260 & 2000jq58 & S & 179.2 & 0.993 & 0.013 & 0.082 & 0.002 & 0.024 & 2.195 & 0.114 & 0.058 & 0.011 & -0.096 & 0.607 & 0.136 \\
34706 & 2001op83 & V & 154.6 & 0.924 & 0.003 & 0.143 & 0.002 & 0.029 & 1.943 & 0.020 & 0.244 & 0.005 & -0.114 & 1.598 & 0.040  \\ \bottomrule

\mbox{B!C = Band I Center; BIIC = Band II Center}

\end{tabular}
\label{tab:band_analysis}
\end{center}
\end{sidewaystable}

%%%%%%%%%%%%%%%%%%%%%%%%%%%%%%%%%%%%%%%%%%%%%%%%%%%%
\subsubsection{The Spectral Analysis Routine for Asteroids, SARA}
\label{subsec:sara}
\begin{figure}[!h]
\centering

\begin{minipage}[b]{0.46\linewidth}
\hspace{-0.225in}\includegraphics[width=2.75in]{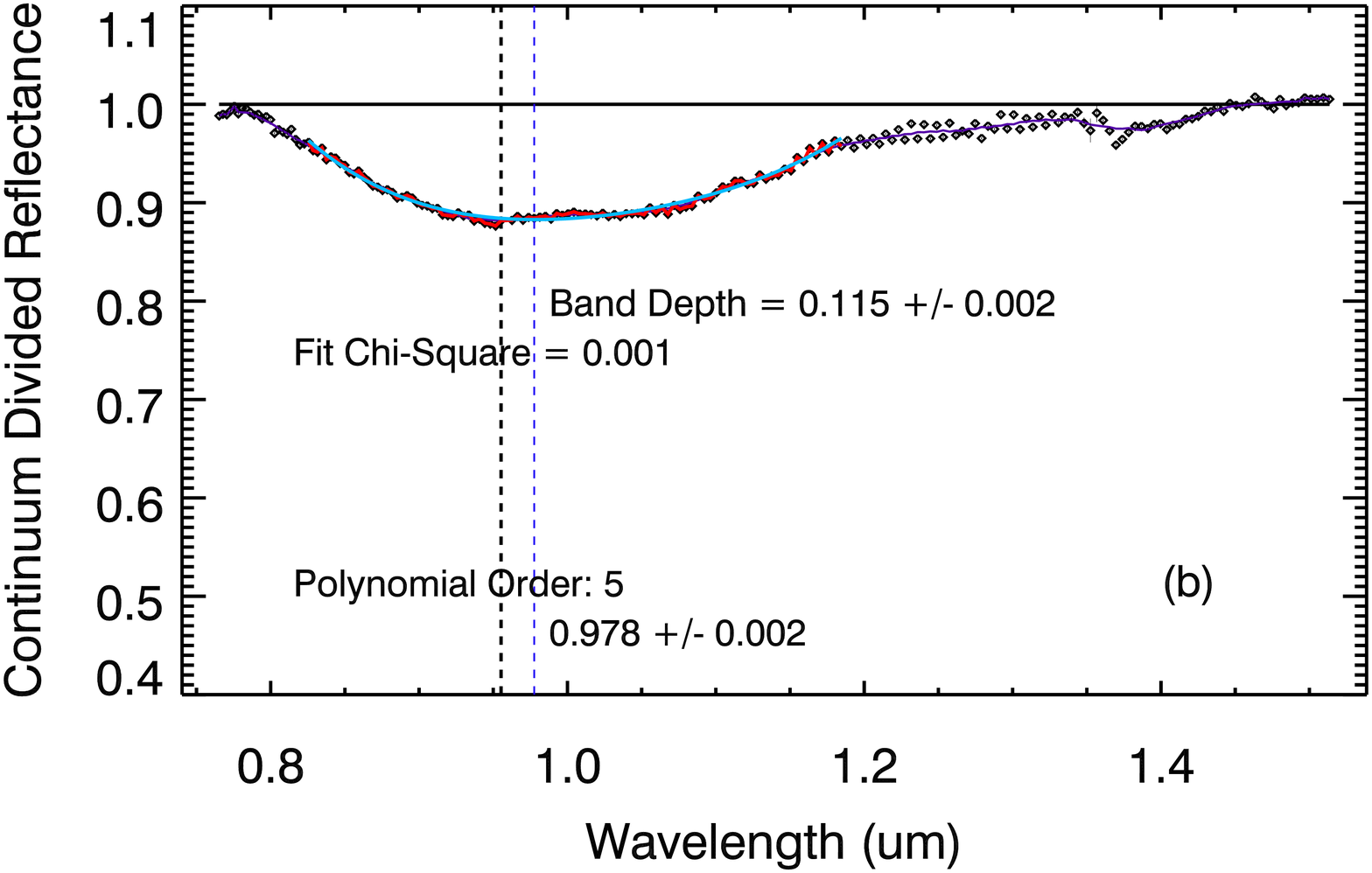}
\end{minipage}
\quad
\begin{minipage}[b]{0.46\linewidth}
\includegraphics[width=2.75in]{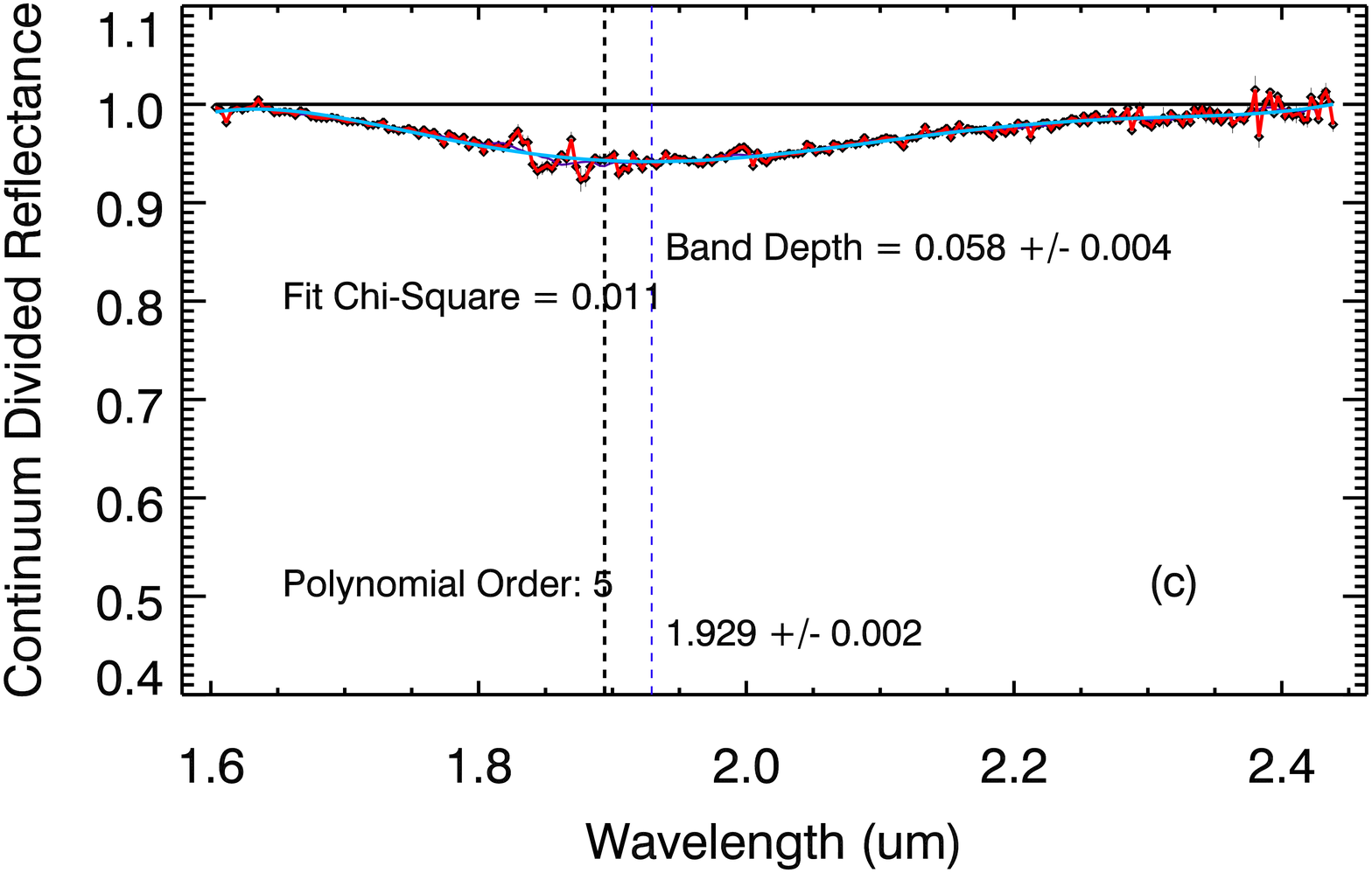}
\end{minipage}
\begin{minipage}[b]{\linewidth}
\hspace{-0.0in}\includegraphics[width=5in]{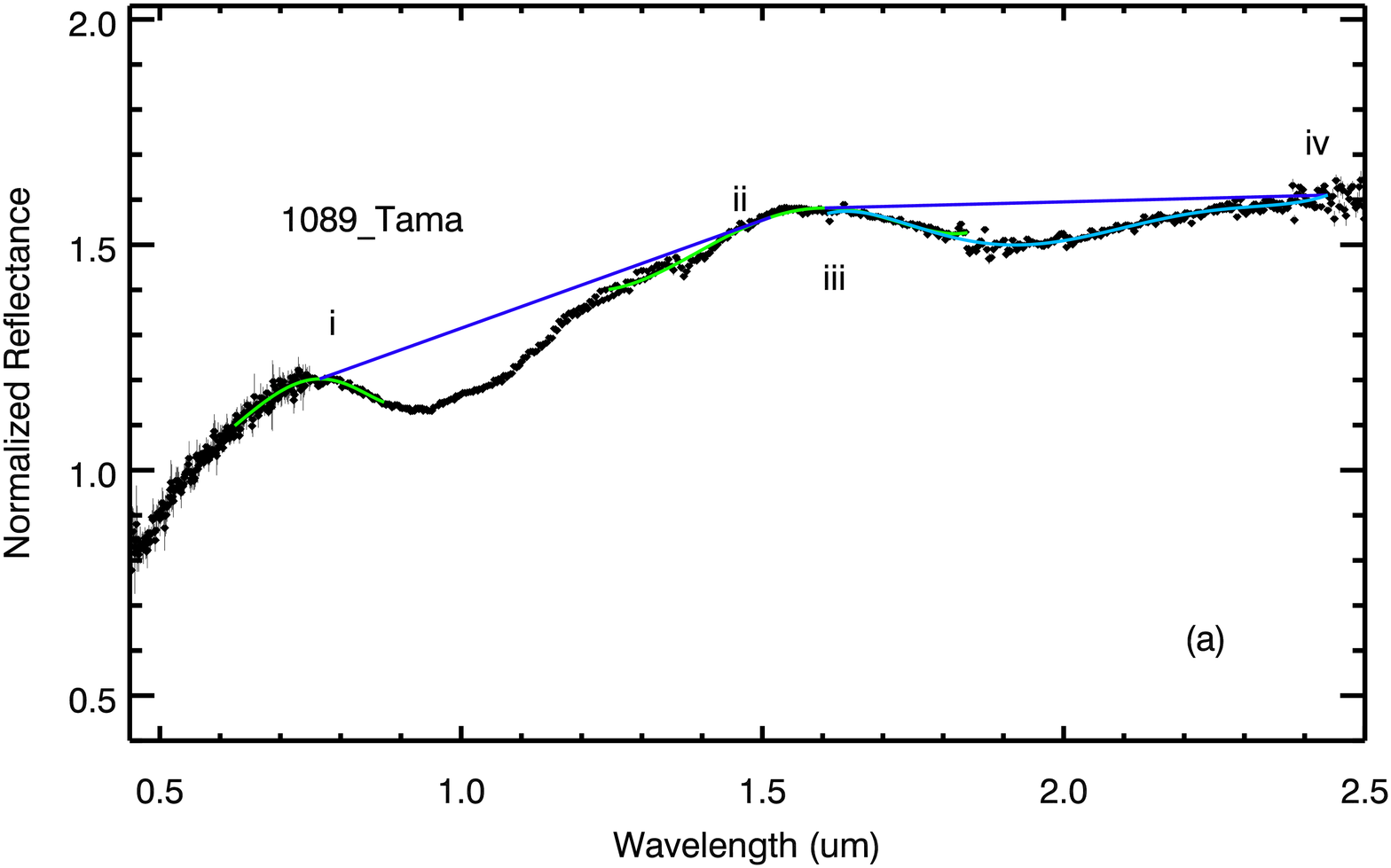}
\end{minipage}
\caption{\emph{(a)} The near-infrared reflectance spectrum of 1089 Tama overlain with the band definitions (i - iv), linear continua (\emph{blue}) for Band I and Band II as well as the $5^{th}$ order polynomial fits to the reflectance maxima near 0.7 and 1.5~\micron~(\emph{green}) and the entirety of Band II (\emph{cyan}). \emph{b} The continuum-divided Band I with data range the polynomial fits \emph{cyan} are applied to highlighted in \emph{red}.  The vertical dashed lines are the minimum of a ten-point boxcar smoothing of Band I (\emph{black}) and the band center (\emph{blue}) determined by the $3^{rd}$,$4^{th}$, and $5^{th}$ order polynomial fits to bottom half of Band I.  \emph{(c)} Same as (b) but for Band II.  For Band II the n$^{th}$ order polynomial fits are applied to the entire band for S-complex asteroids and the bottom half for V-type asteroids.  See text for the definitions of points (i - iv).}
%Points (i and iii) are the maxima to the 5$^{th}$ order polynomial fits to reflectance maxima near 0.75 and 1.5~\micron (\emph{green}).  Point (ii) is the point where a line drawn from point (i) lies tangent to the polynomial fit to the 1.5~\micron \ maximum.  Point (iv) is the 2.44~\micron \ reflectance value of the $5^{th}$ order polynomial fit to Band II (\emph{cyan}). \emph{(b)} Band I divided by the linear continuum.
\label{fig:band_algorithm}
\end{figure}

The first step to measure band parameters is to define boundaries for Band I and Band II.  Figure~\ref{fig:band_algorithm}a indicates the four crucial points (i, ii, iii, and iv) for this determination, where the point pairs (i, ii) and (iii, iv) define the boundaries of Band I and Band II, respectively.  Points (i) and (iii) are defined by the two reflectance maxima near 0.75 and 1.4~\micron.  SARA fits 5$^{th}$ order polynomials to reflectance data range defined by $\pm 0.3$~\micron \ of the data maxima near 0.75 and 1.4~\micron.  The maxima of these fits are then defined to be points (i) and (iii).  Fifth order polynomials are chosen to maximize the accuracy and generality of the algorithm.  These regions can exhibit complicated structure due to Band I being a composite of three olivine and one pyroxene absorption as well as (potentially) poor telluric corrections to the spectral data near 1.35 and 1.9~\micron; trial and error testing demonstrated that lower order polynomials are insufficient in replicating this structure in a repeatable manner.  Point (ii) is defined as the point where a line extending from point (i) lies tangent to the polynomial fit of the 1.4~\micron \ region.  By using the tangent point instead of the point (iii) maximum, the band analysis algorithm ensures the continuum does not undercut reflectance values.  As such, it more accurately measures the Band I and Band II areas.  

The quality of the IRTF SpeX data decreases significantly longward of 2.45~\micron, and for many of the reflectance spectra in our sample, the signal-to-noise drops precipitously at this point.  To guarantee an accurate representation of reflectance values, the red-edge of Band II, point (iv), is set to the reflectance value at 2.44~\micron \ of a 5$^{th}$ order polynomial fit to the entirety of Band II\footnote{To avoid tracing extremely noisy data longward of $\sim$2.4~\micron, certain asteroid spectra required the use of a 3$^{rd}$ order polynomial fit to characterize point (iv).}. With the boundaries of Band I and II defined, linear continua are generated connecting points (i)-(ii) and (iii)-(iv).  The slope of Band I is recorded as the slope of the Band I continuum, and the algorithm divides each band by its respective continuum.  

The band centers are measured by computing a 3$^{rd}$, 4$^{th}$ and 5$^{th}$ order polynomial fit to the bottom of the continuum-divided band (indicated as red points in Fig.~\ref{fig:band_algorithm}b and c).  Here, the procedure for V-types and S-complex asteroids diverge.  For both S-complex and V-type asteroids, the Band I center is measured by computing polynomial fits to the bottom half of the band. For S-types, with shallower Band IIs that are more sensitive to the effects of observational uncertainties and telluric contamination near 1.9~\micron, the center is measured by computing the polynomial fits to the entirety of Band II.  However, for V-types, with deeper Band IIs that are less sensitive to the effects of observational uncertainties, the center is measured by computing polynomial fits to the bottom half of Band II. The band depth is taken to be one minus a ten-point boxcar smooth of the reflectance at the wavelength position of the band center. The band center and depth are then defined as the average of the measured center and depths over the three polynomial fits.  The band areas are calculated as the area between flat continuum and the continuum divided band reflectance data.  The BAR is computed as (Band II Area) / (Band I Area).  

A graphical example of SARA's assignment of points (i - iv) and the linear continua fits is provided in Fig.~\ref{fig:band_algorithm}. Figs. \ref{fig:band_algorithm}b and \ref{fig:band_algorithm}c show the continuum divided Bands I and II with the polynomial fits to determine the band centers.  The band parameters determined by SARA for our 24 S-complex and V-type asteroids are presented in Table~\ref{tab:band_analysis}.
%For low signal-to-noise in Band II, the entire band is required to ensure a robust polynomial fit.  While the error introduced by using whole-band polynomial fits would be a concern for  the asymmetric Band I, it is not a large source of band center error for Band II due to the nearly symmetric quality of Band II. 

%%%%%%%%%%%%%%%%%%%%%%%%%%%%%%%%%%%%%%%%%%%%%%%%%%%%
\subsubsection{Band Parameter Errors}
\label{subsec:ba_code_error}

SARA assesses the 1$\sigma$ error bars using Monte Carlo simulations.  Using Monte Carlo simulations to estimate errors is particularly important for band center errors because the error is in the wavelength direction, but the observational uncertainties are along the reflectance axis.  For Band I and II, gaussian noise set to the level of the measured signal-to-noise is added to the measured reflectance values.  This gives a new set of band reflectance points that are consistent with the observational uncertainties.  New 3$^{rd}$, 4$^{th}$ and 5$^{th}$ order polynomials are fit to the simulated reflectance values and the band centers, depths, and areas are re-measured.  For each polynomial order, this process is iterated 20,000 times to give $3 \times 20,000$ measurements for band center and depth.  For each polynomial order, the band center and depth error is taken to be the standard deviation of the 20,000 measurements, and the final 1$\sigma$ center and depth errors are assigned to be the average of the three error measurements.  The measured band area is independent of the polynomial fits, and the 1$\sigma$ band area error is taken as the standard deviation of the 20,000 measured areas.  The Band I and II center and area errors are listed in Table~\ref{tab:band_analysis}.

SARA also computes Band I slope errors, but as they are not used in this MAS study, they are not listed in Table~\ref{tab:band_analysis}.  This error is also assessed using an iteration of 20,000 Monte Carlo simulations, but here the Monte Carlo simulation is applied to the reflectance values used to define points (i, ii, and iii).  Using simulated reflectance values, new 5$^{th}$ order polynomials are fit to identify points (i) and (iii), which in turn defines the tangent point (ii).  The value of the slope is recorded for each simulation, and the final error is taken to be the standard deviation of the 20,000 measured slopes.  

%%%%%%%%%%%%%%%%%%%%%%%%%%%%%%%%%%%%%%%%%%%%%%%%%%%%
\subsubsection{Band Parameter Temperature Corrections}
\label{subsec:ba_tcor}

The temperature of the asteroid surfaces influences the band center positions and widths of the 1- and 2-\micron \ bands \citep{Singer:1985,Schade:1999,Moroz:2000,Hinrichs:2002,Sanchez:2012}.  Laboratory measurements of band parameters are typically obtained at near room temperature ($\sim$300 K), and will therefore have slightly different band parameters than asteroid surfaces.  Hence, corrections to the band parameters at asteroidal temperatures are required when comparisons or calibrations based on laboratory measurements are made.  \citet{Sanchez:2012} found that Band II is the most affected by temperature effects and derived the following equations to correct for these effects with respect to room temperature:

\begin{eqnarray}
 \Delta BIIC(\mu m) & = & 0.06 - 0.0002 \times T({\rm K}) \\
 \Delta BAR & = & 0.00075 \times T({\rm K}) - 0.23.
\label{eqn:tcor}
\end{eqnarray}

The surface temperatures of the asteroids are estimated using the equilibrium temperature equation, 

\begin{equation}
T_{eq} = \left[ \frac{L_\odot (1 - A)}{16\pi\sigma\epsilon  r^2_h} \right]^\frac{1}{4},
\end{equation}

where $L_\odot$ is the Solar luminosity ($3.846 \times 10^{26}$W), $A$ is the bolometric albedo, $\sigma$ is the Stefan-Boltzmann constant, $\epsilon$ is the bolometric emissivity, and $r_h$ is the heliocentric distance of the asteroid at the time of observation.  For our sample of MASs, we use bolometric albedos from the Small Bodies Planetary Data Systems (PDS) Node\footnote{http://sbn.pds.nasa.gov/}, and the bolometric emissivity is set to be 0.9.  The temperatures, $\Delta BIIC$, and $\Delta BAR$ of the asteroids are listed in Table~\ref{tab:band_analysis}, and hereafter, all listed and utilized BIIC and BAR values are temperature corrected.

%%%%%%%%%%%%%%%%%%%%%%%%%%%%%%%%%%%%%%%%%%%%%%%%%%%%
\subsection{S-subtypes and Meteorite Analogs}
\label{subsec:ba_ssub}

\begin{figure}[htb]
\begin{center}
\includegraphics[width=12cm]{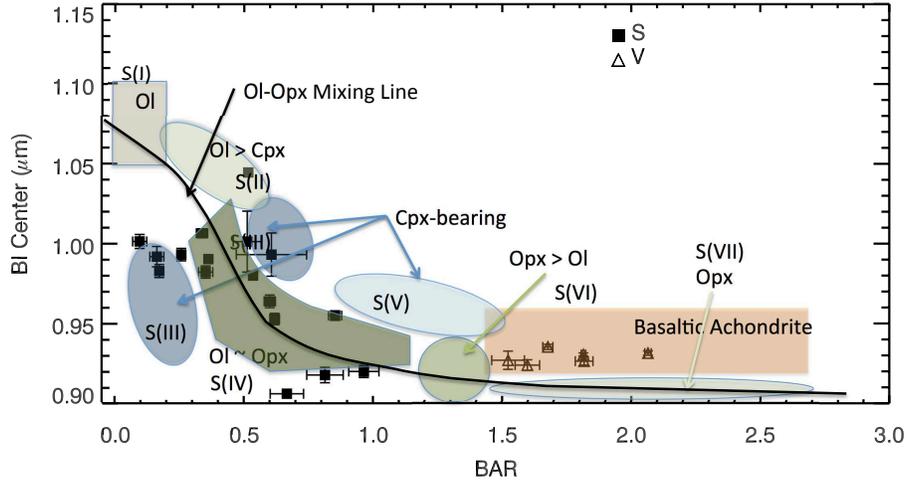}
\caption{The Band I Center versus BAR spectral parameters of 18 S-type (\emph{solid squares}) and 6 V-type (\emph{open triangles}) MASs over-plotted on the S(I) - S(VII) regions plus basaltic achondrite region defined in \citet{Gaffey:1993}.  The solid curve is the olivine-orthopyroxene mixing line defined in \citet{Cloutis:1986}, and the qualitative relative abundance of each zone is listed for each zone.  The 1$\sigma$ uncertainties for Band I Center and BAR in Table~\ref{tab:band_analysis} are plotted for each object.}
\label{fig:gaffey_plot}
\end{center}
\end{figure}

The Band I center and BAR band parameters can be used to parse the S-type asteroids into seven sub-types, S(I - VII) that roughly trace the mineralogical characteristics of the asteroid \citep{Gaffey:1993}).  The Band I centers versus BARs for our S- and V-type MASs are plotted in Fig.~\ref{fig:gaffey_plot}.  Using mixtures of olivine (ol) and orthopyroxne (opx), \citet{Cloutis:1986} demonstrated that a plot of Band I center versus BAR center produces a distinct curve for mixtures ranging from pure ol to pure opx.  Under this same parameterization, the S-type asteroids cluster around the ol-opx mixing line.  This provides the basis that S-type asteroids represent a suite of mineralogies ranging from pure olivine, S(I), to pure pyroxene, S(VII).  Deviations above and below the ol-opx mixing line are caused by shifts in the Band I center to longer wavelengths with increasing amounts of calcium (\Ctwo) within the pyroxene phases \citep{Adams:1974,Gaffey:1993}.  

\begin{figure}[htb]
\begin{center}
\includegraphics[width=12cm]{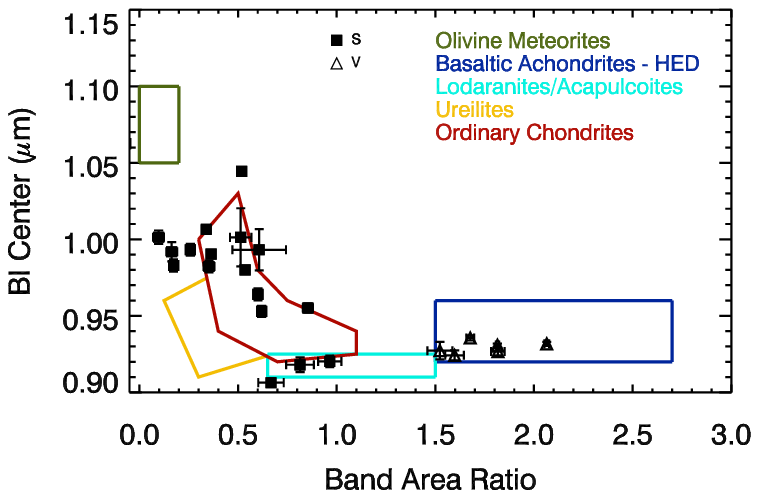}
\caption{The Band I Center versus BAR spectral parameters of 18 S-type (\emph{solid squares}) and 6 V-type (\emph{open triangles}) MASs over-plotted on regions of potential meteorite analogs. The olivine, ordinary chondrite, and basaltic achondrite regions are from \citet{Gaffey:1993}; the primitive achondrite region (lodranites/acapulcoites) is from \citet{Burbine:2001}; and the ureilite region is from \citet{Cloutis:2010}.  The 1$\sigma$ uncertainties for Band I Center and BAR in Table~\ref{tab:band_analysis} are plotted for each object.}
\label{fig:gaffey_analogs}
\end{center}
\end{figure}

The general meteorite analog is determined using a modified version of the S-subtype plot (Fig~\ref{fig:gaffey_analogs}).  In the Band I center vs. BAR parameterization, the subtypes S(I), and S(IV), and the basaltic achondrite (BA) zone above the S(VII) zone, correspond to known meteorite assemblages, such that S(I)s are mono-mineralic, olivine meteorite assemblages (e.g., pallasites and brachinites), S(IV)s are ordinary chondritic (OC) assemblages, and BAs are HED (Howardite, Eucrite, Diogenite) assemblages \citep{Gaffey:1993,Burbine:2007,Burbine:2009,Dunn:2010min,Reddy:2011b}.  Because of the shape of the S(IV)/OC region, this zone is commonly referred to as the `OC Boot'.  Of the three meteorite zones, the S(IV)s and BAs have positively been linked to specific meteorite groups, and hence, provide the most reliable means of identifying a specific meteorite analog using band parameter analyses results (see \S~\ref{sec:oc_min}, \ref{sec:vtype_min}). 

A summary of the S-subtypes with likely meteorite analogs for the MAS sample is provided in Table~\ref{tab:mineralogy}. Several of the MASs fall near or on a S-subtype boundary in Fig.~\ref{fig:gaffey_plot}.  For these objects, both subtypes are listed in Table~\ref{tab:mineralogy}.  In total, we classify 1 S(I), 1 S(II), 3 S(III), 3 S(III)/S(IV) border, 7 S(IV), and 3 sub-S(IV) object out of the 18 S-type MASs in the sample.  The sub-(IV) classification is given to 2131 Mayall, 3623 Chaplin, and 5407 (1992 AX), which fall below the `OC Boot'. 3623 Chaplin and 5407 (1992 AX) plot on the S(IV) zone lower boundary, and therefore, are likely to be analogous to OCs.  However, 2131 Mayall plots well below the S(IV) zone in a region that may be associated with primitive achondrites (lodranites and acapulcoites), and/or the anomalous ureilite meteorite group \citep{Burbine:2001,Cloutis:2010,Lucas:2014}.

\begin{table} \scriptsize \renewcommand{\arraystretch}{0.8} \vspace{0.005in}
\caption{S-subtype, Mineralogy, and Potential Meteorite Analog}
\begin{center}
\begin{tabular}{ p{0.6cm} p{1.25cm} p{0.6cm} p{1.25cm} p{0.5cm}  p{0.7cm} p{0.7cm}  p{0.8cm} p{1.90cm} } 

\toprule
ID & Object & BDM & S-sub & MAS & Fs & Fa & ol/ & Meteorite \\ 
 &  &  &  & Type & (mol\%) & (mol\%) & (ol+opx) & Analog \\ \toprule
939 & Isberga & S & S(III) & U & 24.19 & 29.42 & 0.70 & OC (LL)? \\
1089 & Tama & S & S(IV) & T2 & 22.43 & 27.07 & 0.60 & L/LL-Chondrite \\
1313 & Berna & Sq Q & S(IV) & T2 & 23.37 & 28.34 & 0.64 & LL-Chondrite \\
1333 & Cevenola & Sr & S(III)/S(IV) & T3 & 24.19 & 29.41 & 0.60 & LL-Chondrite? \\
1509 & Esclangona & S & S(IV) & T3 & 19.02 & 22.38 & 0.58 & L-Chondrite \\
1717 & Arlon & S & S(III) & T3 & 22.71 & 27.45 & 0.69 & OC (LL)? \\
2131 & Mayall & S & sub-S(IV) & T3 & 10.14 & 9.91 & 0.57 & Lodranite/ \\
& & & & & & & & Acapulcoite? \\
2577 & Litva & S & S(II) & T3 & 25.33 & 30.62 & 0.60 & ? \\
3309 & Brorfelde & S & S(IV) & T3 & 19.34 & 22.82 & 0.52 & L-Chondrite \\
3623 & Chaplin & Sq Q & sub-S(IV) & T4 & 12.73 & 13.56 & 0.53 & H-Chondrite/ \\
& & & & & & & & Los.-Aca. \\
3749 & Balam & Sq Q & S(IV) & T3 & 22.64 & 27.36 & 0.64 & LL-Chondrite \\
4674 & Pauling & S & S(IV) & T3 & 24.5 & 29.81 & 0.65 & LL-Chondrite \\
5407 & 1992ax & S & sub-S(IV) & T3 & 13.18 & 14.21 & 0.58 & H-Chondrite/ \\
& & & & & & & & Ureilite \\
5905 & Johnson & Sq Q & S(IV) & T3 & 20.56 & 24.52 & 0.58 & L-Chondrite \\
6265 & 1985tw3 & Sq Q & S(III)/S(IV) & T3 & 23.6 & 28.65 & 0.67 & LL-Chondrite \\
7225 & Huntress & S & S(I) & T3 & 24.83 & 29.69 & 0.73 & Olivine \\
8116 & Jeanperrin & Sq Q & S(III) & T3 & 23.5 & 28.51 & 0.69 & OC (LL)? \\
17260 & 2000jq58 & S & S(III)/S(IV) & T3 & 23.6 & 28.65 & 0.58 & LL-Chondrite \\ \bottomrule

\mbox{BDM is Bus-DeMeo Taxa ~~~~~U is Unclassified MAS Type}
\mbox{Analogs with ? indicate uncertain meteorite connections}

\end{tabular}
\label{tab:mineralogy}
\end{center}
\end{table}

%%%%%%%%%%%%%%%%%%%%%%%%%%%%%%%%%%%%%%%%%%%%%%%%%%%%
%\subsection{Mineralogical analyses}
%\label{sec:ba_mineralogy}

%Spectral band parameter analyses have been used to determine appropriate meteorite analogs and infer asteroid mineralogy since \citet{Gaffey:1993} first established the S-subtypes and connected the S(IV), BA, and S(I) zones in Band I Center vs BAR space to ordinary chondrites, basaltic achondrites, and olivine meteorites, respectively.

%introduce the use of band parameter analyses to infer asteroid mineralogy and determine appropriate meteorite analogs.  Highlight the laboratory calibrations for OC, HED, and olivine meteorites.
%Such calibrations exists for S(IV) \citep{Dunn:2010}, S(I) \citep{King:1987,Reddy:2011b}, and V-type \citep{Burbine:2007,Burbine:2008}
%The band parameter analysis can be used to determine the mineralogy for S(IV), S(I), and V-type asteroids through the application of calibration equations determined by thorough laboratory analyses of meteorite analogs.

%Band parameter analyses have been used to determine mineralogical information of asteroids since \citet{Gaffey:1993} characterized spectral parameters of ordinary chondrites, basaltic achondrites, and olivine meteorites.  

%%%%%%%%%%%%%%%%%%%
\subsubsection{S(IV) Mineralogy}
\label{sec:oc_min}

%The band parameter analysis can be utilized to determine mineralogical information.  

The S(IV) subtype has been connected to the OC meteorites since \citet{Gaffey:1993}.  However, due to the multiple phases of pyroxene in OCs \citep{Gaffey:2007}, accurate calibrations connecting the spectral band parameters to mineralogy did not exist until the \citet{Dunn:2010P1,Dunn:2010min,Dunn:2010P2} series of studies, hereafter Dunn 2010 Series, improved upon the normative abundance calibrations of \cite{Burbine:2003}.  Using a sample of 48 OCs, the Dunn 2010 Series connects X-ray diffraction measured relative abundances, the ol ratio, and mol\% Fs and Fa in pyroxene and olivine, respectively, to a 1- and 2-\micron \ band parameter analysis from IR spectral data.   A major result of the Dunn 2010 Series is a set of calibrations to equate Band I centers (BIC) to an ol ratio (Eqn.~\ref{eqn:ol_ratio}) and BAR to modal abundances (Eqns.~\ref{eqn:fs} and \ref{eqn:fa}).  The mineralogical results of applying the S(IV), S(IV)/S(III), and sub-S(IV) MAS band parameters to these equations are shown in Table~\ref{tab:mineralogy}. 

%These relative abundance results are calibrated to measured modal abundances, Fs in pyroxene and Fa in olivine \citep{Brearley:1998}, for of each of the OC groups (H-, L-, and LL-chondrites).  These laboratory measurements were calibrated to the BAR and Band I Centers measured from the IR spectra of the same 48 OCs to determine the ol ratio (Eqn~\ref{eqn:ol_ratio}) and modal abundances (Eqns~\ref{eqn:fs} and \ref{eqn:fa}).
%The mineral compositions and abundances for asteroids with determined NIR band parameters can be determined in cases where laboratory spectral calibrations have been performed.  Using the meteorite analogs of ordinary chondrites, \citet{Dunn:2010min} determined band parameter calibrations to determine relative abundance, ol/(ol+opx) (Eqn~\ref{eqn:ol_ratio}) from BAR measurements.   

\begin{equation}
ol/(ol + opx) = -0.242 \times BAR + 0.728
\label{eqn:ol_ratio}
\end{equation}

%The \citet{Dunn:2010min} experiment also provides calibrations for mineral composition, mol\% ferrosilite (Fs) in pyroxene and mol\% fayalite (Fa) in olivine (Eqns.~\ref{eqn:fs} and \ref{eqn:fa}) from Band I center (BIC) measurements.

\begin{equation}
Fs = -879.1 \times~ (BIC)^2 + 1824.9 \times (BIC) - 921.7
\label{eqn:fs}
\end{equation}

\begin{equation}
Fa = -1284.9 \times (BIC)^2 + 2656.5 \times (BIC) - 1342.3
\label{eqn:fa}
\end{equation}

\begin{figure}[htb]
\begin{center}
\includegraphics[width=10cm]{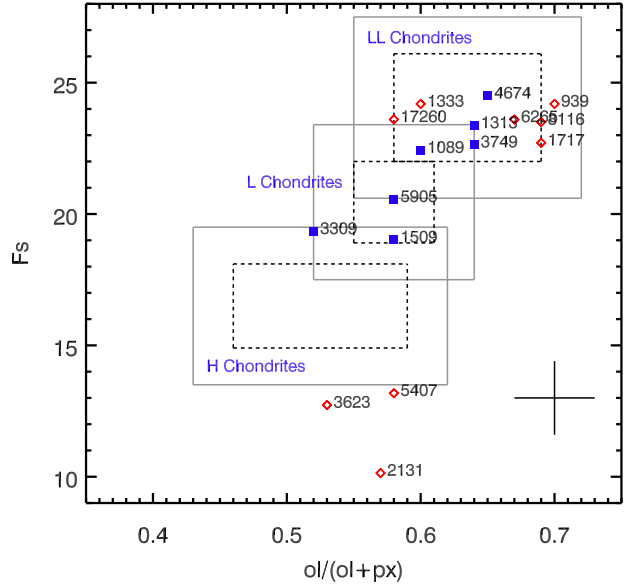}
\caption{The mol\% Fs in pyroxenes as a function of ol/(ol + opx) for the S(IV), S(IV)/S(III), and sub-S(IV) MASs.  The dashed boxes represent the range of ol/(ol + opx) measured via X-ray diffraction by the Dunn 2010 Series and the range of Fs contents in ordinary chondrites \citep{Brearley:1998} for the H, L, and LL chondrites.  The \emph{gray} solid boxes include the uncertainties in the calibration equations from the Dunn 2010 Series (+/- 0.03 for ol/(ol + opx) and 1.4 mol\% for Fs).  The \emph{blue squares} are MASs that cleanly fall within the S(IV) region in Fig.~\ref{fig:gaffey_plot}.  The \emph{red open diamonds} are MASs that fall on the border of the S(IV) and another zone (either S(III) or S(V)), or in the case of 2131, fall well below the S(IV) zone. The crossbar designates the aforementioned errors in the calibrations.}
\label{fig:fs}
\end{center}
\end{figure}

The mineralogy results can be employed to further narrow down the appropriate meteorite analog to the H, L, and LL chondrite OC subgroups.  These three subgroups are based on their relative amount of metallic iron content.  The H-chondrites are the highest in metallic iron content, the L chondrites are low in metallic iron content, and the LL chondrites are the lowest in metallic iron content.  Conversely, this translates to the inverse relationship with respect to FeO content, such that the H chondrites are lowest in mol\% Fs/Fa and LL chondrites are highest in mol\% Fs or Fa.  Additionally, the `OC Boot' has a strong ol ratio gradient that is reflective of the variations in olivine-to-pyroxene in the OC subgroups.  These trends in relative and modal abundance map to three defined regions in ol ratio vs. Fs or Fa space based upon the modal and relative results and uncertainties computed from the OCs (Dunn 2010 Series).  We compare the mineralogical results for the MASs that are classified as S(IV) with the H, L , and LL chondrite ol ratio vs. Fs zone in Fig.~\ref{fig:fs}.  Individual error bars are not included for the MASs because the uncertainty in the calibration equations, +/- 0.03 ol/(ol + opx) and +/- 1.4 mol\% Fs, is larger than the propagated errors from the band parameter analysis.  The H, L, and LL subtype results for ol ratio vs. Fa zone are the same as the ol ratio vs. Fs zone, and as such, are not shown.
%We compare the mineralogical results for the MASs that are classified as S(IV) with the H-, L-, and LL-chondrite ol ratio vs. Fs or Fa zones in Figs.~\ref{fig:fs} and \ref{fig:fa}, respectively.  Individual error bars are not included for the MASs because the uncertainty in the calibration equations, +/- 0.03 ol/(ol + opx) and +/- 1.3 (1.4) mol\% Fa (Fs), is larger than the propagated errors from the band parameter analysis.

%\begin{figure}[htb]
%\begin{center}
%\includegraphics[height=4in,width=5in]{fig07.eps}
%\caption{The mol\% Fa in olivine's as a function of ol/(ol + opx) for the S(IV), S(IV)/S(III), and sub-S(IV) MASs.  The dashed boxes represent the range of ol/(ol + opx) measured via X-ray diffraction by the Dunn 2010 Series and the range of Fa contents in ordinary chondrites \citep{Brearley:1998} for the H-, L-, and LL-chondrites.  The \emph{gray} solid boxes include the uncertainties in the calibration equations from the Dunn 2010 Series (+/- 0.03 for ol/(ol + opx) and 1.4 mol\% for Fs).  The \emph{blue squares} are MASs that cleanly fall within the S(IV) region in Fig.~\ref{fig:gaffey_plot}.  The \emph{red open diamonds} are MASs that fall on the border of the S(IV) and another zone (either S(III) or S(V)), or in the case of 2131, fall well below the S(IV) zone. The crossbar designates the aforementioned errors in the calibrations.}
%\label{fig:fa}
%\end{center}
%\end{figure}

Out of the seven MASs (\emph{solid blue squares} in Fig.~\ref{fig:fs}) in Table~\ref{tab:mineralogy} that cleanly fall into the `OC Boot', three (1313 Berna, 3749 Balam, and 4674 Pauling) are analogous to the LL chondrites; three (1509 Esclangona, 3309 Brorfelde, and 5905 Johnson) are analogous to the L chondrites; and one (1089 Tama) lies between the L and LL chondrite zones.  Applying the calibration equations to include the S(IV)/S(III) border, S(III)s, and sub-S(IV)s regions close to the S(IV) `OC Boot' (\emph{open red triangles} in Fig.~\ref{fig:fs}, expands the number of MASs included in the analysis to 16.  Assuming the Dunn 2010 Series calibrations are accurate for this expanded range, there are two H-Chondrite , three L-chondrite, nine LL-chondrite, and one L/LL-chondrite analog.  Again, 2131 Mayall is anomalous, and it may be related to the primitive achondrites - lodranites and acapulcoites \citep{Burbine:2001} or ureilites \citep{Cloutis:2012}.  This gives 15 MASs with OC-like compositions, where 2 (13.33\%) are H-like, 3 (20.00\%) are LL-like, 1 (6.67\%) is L/LL-like, and 9 (60.00\%) are LL-like.  Compared with meteorite falls, where LL chondrites are rare, this is a significant excess of LL-like mineralogies.  The potential implications of these statistics compared with the meteorite population and the near-Earth asteroid (NEA) population are discussed in \S\ref{sec:too_many_LLs}.

%%%%%%%%%%%%%%%%%%%
\subsubsection{V-type Mineralogy}
\label{sec:vtype_min}

The pyroxene modal abundances for the V-type MASs are computed using calibration equations (Eqns~\ref{eqn:hed_fs} and \ref{eqn:hed_wo}) derived from comparing HED Band I and II centers to mineralogy determined via microprobe analysis \citep{Burbine:2007, Burbine:2009}.  

\begin{eqnarray}
 Fs(\pm 3~mol\%) &=& 1023.4 \times (BIC) - 913.82 \nonumber \\
 Fs(\pm 3~mol\%) &=& 205.86 \times (BIIC) - 364.3 
\label{eqn:hed_fs}
\end{eqnarray}

\begin{eqnarray}
 Wo(\pm 1~mol\%) &=& 396.13 \times (BIC) - 360.33 \nonumber \\
 Wo(\pm 1~mol\%) &=& 79.905 \times (BIIC) - 148.3,
\label{eqn:hed_wo}
\end{eqnarray}

The final modal abundances are taken as the average of the Band I and Band II center results, and the enstatite abundance is taken as 1 - (Fs + Wo).  The results of applying these calibration equations to our six V-type MASs are summarized in Table~\ref{tab:vtype_min}.

\begin{figure}[htb]
\begin{center}
\includegraphics[width=10cm]{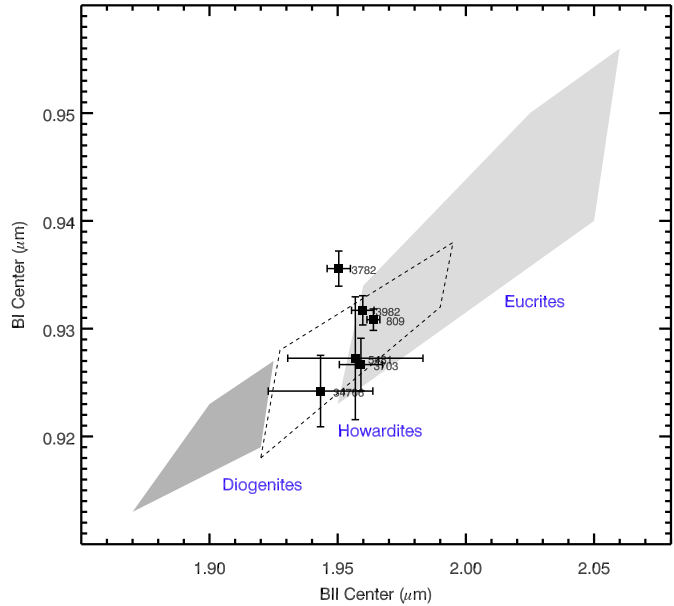}
\caption{The measured band centers of V-type MASs with their derived 1~$\sigma$ error bars overplotted on the Howardite, Eucrite, and Diogenite zones determined in \citet{Moskovitz:2010}.}
\label{fig:hed_b2vb1}
\end{center}
\end{figure}

We attempt to identify the specific HED meteorite analog (howardite, eucrite, or diogenite) for the V-type MASs following the analysis of \citet{Moskovitz:2010}.  In \citet{Moskovitz:2010}, howardite, eurcrite, and diogenite regions are defined for Band I versus Band II center parameter space using a sample 75 HED meteorites.  We compare the V-type MASs band parameters to these region in Fig.~\ref{fig:hed_b2vb1}.  All six of the V-type MASs fall within the overlap of the howardite/eucrite regions preventing any positive relation with a specific HED meteorite analog for the V-type MASs.  However, in agreement with the observed rarity of diogenite-like spectra, none of the V-type MASs match the diogenite region. 

\begin{table} \scriptsize \renewcommand{\arraystretch}{0.8} \vspace{0.005in}
\caption{Derived mineralogy of V-type MASs}
\begin{center}
\begin{tabular}{ p{0.6cm} p{1.7cm} p{0.6cm} p{0.9cm} p{0.9cm} p{0.9cm} p{0.9cm}} 

\toprule
ID & Object & BDM & MAS Type & Fs (mol\%) & Wo (mol\%) & En (mol\%) \\ \toprule
809 & Lundia & V & T2 & 39.39 & 8.40 & 52.20 \\
3703 & Volkonskaya & V & T3 & 36.76 & 7.38 & 55.85 \\
3782 & Celle & V & T3 & 40.43 & 8.80 & 50.77 \\
3982 & Kastel & V & T3 & 39.41 & 8.41 & 52.18 \\
5481 & Kiuchi & V & T3 & 36.84 & 7.41 & 55.75 \\
34706 & 2001op83 & V & T3 & 33.88 & 6.27 & 59.85   \\ \bottomrule

%\mbox{BDM is Bus-DeMeo Taxa ~~~~~U is Unclassified MAS Type}
%\mbox{Analogs with ? indicate uncertain meteorite connections}

\end{tabular}
\label{tab:vtype_min}
\end{center}
\end{table}

%%%%%%%%%%%%%%%%%%%
%\subsubsection{S(I) Mineralogy}
%\label{sec:si_min}

%%%%%%%%%%%%%%%%%%%%%%%%%%%%%%%%%%%%%%%%%%%%%%%%%%%%%%
%%%%%%%%%%%%%%%%%%%%%%%%%%%%%%%%%%%%%%%%%%%%%%%%%%%%%%
\section{Internal Structure of MAS sample}
\label{sec:macroporosity}

A primary goal of this study is to use the MAS taxonomy and mineralogy results to link MASs to an appropriate meteorite analog and apply that connection to the MAS bulk density measurements in \citet{Marchis:2012}.  The analog-density connection makes it possible to characterize the porosity of the MAS.  The porosity is a measure of the internal structure of the MAS, which provides a critical constraint  to understanding formation and evolutionary pathways (see \S\ref{sec:intro}).  

The bulk porosity, $P_b$, or total percentage vacuum space within a MAS, can be calculated by comparing the MAS system density to that of its meteorite analog (Eqn.~\ref{eqn:bulk_porosity}).

\begin{equation}
P_b = \left[ 1 - \frac{\rho_{MAS}}{\rho_{grain}} \right] \times 100,
\label{eqn:bulk_porosity}
\end{equation}

where $\rho_{MAS}$ is the bulk density of MAS, and $\rho_{grain}$ is the grain density of the meteorite analog. The bulk porosity also includes the vacuum associated with the microscopic pore spaces within the asteroidal material.  The pore space vacuum percentage, or microporosity, is assumed to be the same as the microporosity of the associate meteorite analog and is calculated as

\begin{equation}
P_{met.} = \left[ 1 - \frac{\rho_{b,met.}}{\rho_{grain}}\right] \times 100,
\label{eqn:microporosity}
\end{equation}

where $\rho_{b,met.}$ and $\rho_{grain}$ are the bulk and and grain density of the meteorite analog, respectively. The microporosity needs to be subtracted from the bulk porosity to measure the large-scale internal structure; this value is the macroporosity ($P_m = P_b - P_{met.}$), and it is the factor we use to evaluate the MAS formation hypotheses outlined in \S\ref{sec:intro}.

The calculated bulk- and macro- porosities for the MASs with density measurements from \citet{Marchis:2012} are shown in Table~\ref{tab:porosities}.  The meteorite analog densities and micro-porosities are taken from \citet{Consolmagno:2008}.  The uncertainties in bulk and macroporosity are propagated using the bulk density uncertainties for the MASs from \citet{Marchis:2012} and the uncertainties for the meteorite grain density and microporosity from \citet{Consolmagno:2008}.  An average density for the HED meteorites is applied for the V-type MASs since no specific connection to howardite, eucrite, or diogenite is made (see \S~\ref{sec:vtype_min}).  Meteorite analogs followed by a `?' in Table~\ref{tab:porosities} indicate a possible, but unconfirmed, analog.  Both 121 Hermione and 130 Elektra are associated with CM chondrites due to the presence of an absorption near 0.7~\micron \ \citep{Cloutis:2011a,Cloutis:2011b,Cloutis:2012}.  283 Emma is also associated with CM chondrites \citep{Fornasier:2011} despite lacking the 0.7~\micron \ feature, which is common, but not ubiquitous in CM chondrites \citep{Cloutis:2011b}.  The association between 939 Isberga and ureilite is tentative and made because of its location in Fig.~\ref{fig:gaffey_analogs} far to left of the OC boot and above the \citet{Cloutis:2010} ureilite region.

\begin{sidewaystable} \scriptsize \renewcommand{\arraystretch}{0.8} \vspace{0.005in}
\caption{MAS Porosities}
\begin{center}
\begin{tabular}{ p{0.5cm} p{1.1cm} p{0.6cm} p{0.6cm} p{0.85cm} p{0.85cm} p{0.85cm}  p{0.85cm} p{1.0cm} p{0.65cm} p{0.65cm}  p{0.65cm} p{0.65cm} } 

\toprule
ID & Asteroid & BDM & MAS & $D_{\rm eff}$ & $3~\sigma$ & Density  & Density & Meteorite & P$_b$ & P$_b$ & P$_m$ & P$_m$ \\
& & & & (km) & $D_{\rm eff}$ & (g/cm$^{3}$) & Error & Analog & (\%) & Error & (\%) & Error \\ \toprule
45 & Eugenia & C$^1$ & T1 &  210.0& 31.1 & 1.17 & 0.17 & CM & 59.66 & 5.97 & 36.56 & 10.88 \\
     &               &		& &    &      &         &         & (CI) & (52.44) & (6.95) & (17.44) & (7.14)      \\   
107 & Camilla & L & T1 & 241.6 & 35.0 &1.52 & 0.22 & CO & 55.43 & 7.12 & 44.63 & 7.32\\
121 & Hermione & Ch & T1 & 165.8 & 22.1 & 2.11 & 0.28 & CM & 27.24 & 9.86 & 4.14 & 13.42\\
130 & Elektra & Cgh & T1 & 174.9 & 25.5 & 2.34 & 0.34 & CM & 19.31 & 11.93 & -- & -- \\
216 & Kleopatra & Xe & T1 & 152.5 & 21.3 & 2.50 & 0.35 & Aub./EH & 31.51 & 9.68 & 26.30 & 10.78 \\
283 & Emma & Xk & T1& 143.9 & 19.2 & 0.88 & 0.12 & CM? & 69.66 & 4.22 & 46.56 & 10.03 \\
762 & Pulcova & C & T1 & 149.4 & 18.7 & 0.80 & 0.10 & CM & 72.41 & 3.53 & 49.31 & 9.76 \\
809 & Lundia & V & T2 & 9.6 & 1.1 & 1.77 & 0.20 & HED & 46.15 & 6.35 & 38.44 & 8.35 \\
939 & Isberga & Sw & ? & 11$^a$ & 0.1$^a$ & 1.70 & -- & Ureilite? & 49.10 & -- & 44.01 & -- \\
1089 & Tama & Sw & T2 & 12.2 & 1.6 & 2.50 & 0.40 & L & 29.38 & 11.59 & 21.18 & 12.51 \\
     &               &		 & &   &      &         &         & (LL) & (25.15) & (12.14) & (20.06) & (13.33)  \\
1313 & Berna & Q Sq & T2 & 13.3 & 1.4 & 1.30 & 0.15 & LL & 63.28 & 4.45 & 55.08 & 7.07 \\
3749 & Balam & S/Sq & W$^\dagger$ & 4.7 & 0.5 & 2.50$^b$ & 1.2$^b$ & LL & 29.38 & 34.00 & 21.18 & 34.44 \\
3782 & Celle & Vw & T3 & 6.6 & 0.7 & 2.40 & 1.10 & HED & 26.98 & 33.56 & 19.20   & 33.99 \\ \bottomrule

\mbox{Effective Diameter, $D_{eff}$ and Density citation:  \citep{Marchis:2012}}
\mbox{$\dagger$ Wide binary system under \citet{Pravec:2007} nomenclature.}
\mbox{(a) PDS Small Bodies Node, Binary Minor Planets V.6.0 \citep{Johnston:2013}}
\mbox{(b) Only density range (1.3 - 3.7 g/cm$^3$) given in \citet{Marchis:2012}, }
\mbox{\hspace{0.5cm}so density is taken as midpoint and error set to span the range}

\end{tabular}
\label{tab:porosities}
\end{center}
\end{sidewaystable}

Following the internal structure divisions of \cite{Britt:2002}, i.e., coherent body ($P_m \lesssim 15\%$), fractured body ($15\% \lesssim P_m \lesssim 30\%$), and rubble-pile body ($P_m \gtrsim 30\%$), the observed MASs are primarily `rubble-piles' (7/13), secondarily fractured/rubble bodies (2/13), and lastly coherent bodies (1/13).  The other MAS macroporosity estimates (3/13) are unreliable due to their large errors.  Structurally, the so-called `rubble-piles' refer to asteroids that are a compilation of boulders with a distribution of sizes that are held together through self-gravitation \citep{Michel:2001,Richardson:2005,Tanga:2009,Jacobson:2011rot}.  Our macroporosity findings are consistent with the formation hypotheses expectations, i.e., T2, T3, and T4s have high internal porosity and T1s are the only expected coherent bodies.  All of our T2 and T3 MASs are determined to have a high internal porosity, and likely are rubble-piles; and T1 MASs are the only MAS type with macroporosities consistent with coherent structures.

There are several ambiguities to determining the internal structure for the MASs, but they are not without explanation.  For one, the \citep{Britt:2002} structural divisions are not firmly defined, so MASs that have porosities near the $P_m$ divisions should not firmly be considered within that structural group.  This indicates that 216 Kleopatra and 1089 Tama may be rubble-pile structures.  Additionally, some MASs have more than one likely meteorite analog.  For example, if the analog to 45 Eugenia is a CM chondrite, then it likely is a rubble pile structure, but if the analog is a CI chondrite, then it could be a fractured body.  Moreover, several of the MAS internal structures are indeterminable based upon the porosity errors, i.e., 130 Elektra, 3749 Balam, and 3782 Celle.  130 Elektra has a low bulk porosity ($P_b = 19.31 \pm 11.93$) for a C-type asteroid, such that the microporosity of its analog, CM chondrite \citep[$P_{met} = 23.10 \pm 4.70$;][]{Consolmagno:2008}, is larger than the bulk porosity; this likely indicates that 130 Elektra is a coherent body. 

\section{Discussion}
\label{sec:disc}

%%%%%%%%%%%%%%%%%%%%%%%%%%%%%%%%%%%%%%%%%%%%%%%%%%%%%%
%%%%%%%%%%%%%%%%%%%%%%%%%%%%%%%%%%%%%%%%%%%%%%%%%%%%%%
\subsection{5407 (1992 AX)}
\label{sec:1992ax}

We find 5407 1992 AX to be an S(IV) MAS (Band I Center, BAR $= 0.920 \pm 0.004$~\micron, $0.965 \pm 0.058$) commensurate with its Bus-DeMeo S-type taxonomy.  However, using a similar band parameter analysis method, \citet{Sanchez:2013} find 5407 to fall in the basaltic achondrite assemblage region with a (Band I Center, BAR $= 0.93$~\micron \ $\pm 0.01, 2.31 \pm 0.06$), leading to the conclusion that the parent body of 5407 experienced temperatures high enough to cause at least partial melting.  This associates 5407 with a basaltic, non-cumulate eucrite meteorite analog.  This is in stark contrast with our finding of 5407 as an S(IV) that is most closely associated with an H chondrite analog.   

In an attempt to resolve the significant difference between the two studies, we use SARA to calculate the band parameters of 5407 using the IRTF spectrum from \citet{Sanchez:2013}.  %A comparison between the \citet{Sanchez:2013} 5407 spectrum and our own is displayed in Fig.~\ref{fig:5407}.  
While similar, the \citet{Sanchez:2013} spectrum has a deeper Band II with a negative slope.  Moreover, the reflectance maxima near 1.5~\micron \ between the two spectra are significantly different; SARA returns a point (iii) of 1.552~\micron \ for the \citet{Sanchez:2013} spectrum and a value of 1.470~\micron \ for the SOAR + IRTF spectrum of this study.  These spectral differences give a larger Band II area value for the \citet{Sanchez:2013} 5407 spectrum, such that the calculated, non-temperature-corrected, BARs are 2.04 \citep{Sanchez:2013} and 1.052 (this study).  The SARA BAR value for the \citet{Sanchez:2013} spectrum is still consistent with a basaltic achondrite analog.  A likely cause for the larger BAR value is poor observing quality (V. Reddy, \emph{personal communication}).   As our Bus-DeMeo classification, S-subtype, and mineralogical analysis are all in agreement, we find it much more likely that 5407 is associated with an H-chondrite parent body rather than a basaltic achondrite parent body.  We cannot, however, rule out the possibility of surface heterogeneity.

%\begin{figure}[!h]
%\begin{center}
%\includegraphics[width=8cm]{fig09.eps}
%\caption{The SOAR + IRTF spectrum of 5407 (1992 AX) (\emph{black}) compared to that of \citet{Sanchez:2013} (\emph{blue}).  The primary differences are the location of the reflectance maximum near 1.5~\micron and the depth of Band II.  The \citet{Sanchez:2013} spectrum was obtained under poor weather conditions [V. Reddy, personal communication], which may explain the possible differences; however, we cannot rule out the possibility of surface heterogeneity.}
%\label{fig:5407}
%\end{center}
%\end{figure}

%Run band analysis and rewrite
%A visual inspection of the NIR spectra from \citet[cf. Fig. 5][]{Sanchez:2013} and this work reveals remarkable similarity between the spectra.  The cause for the discrepancy between the two studies is therefore unknown, but likely associated with the band parameter analysis methodologies.  As our Bus-DeMeo classification, S-subtype, and mineralogical analysis are all in agreement, we find it much more likely that 5407 is associated with an H-chondrite parent body rather than a basaltic achondrite parent body.

%%%%%%%%%%%%%%%%%%%%%%%%%%%%%%%%%%%%%%%%%%%%%%%%%%%%%%
%%%%%%%%%%%%%%%%%%%%%%%%%%%%%%%%%%%%%%%%%%%%%%%%%%%%%%
%\subsection{Band Analysis: Additional Sources of Error}
%\label{sec:disc_ba_err}

%%%%%%%%%%%%%%%%%%%%%%%%%%%%%%%%%%%%%%%%%%%%%%%%%%%%%%
\subsection{The Band II red-edge problem}
\label{sec:red_edge_problem}

Due to observational limitations, the end of Band II is not typically observed.  This makes the red-edge of Band II a choice rather than an empirical measurement.  For example, \citet{Cloutis:1986} define point (iv) to be the reflectance at 2.44~\micron, while \citet{Gaffey:2002} define it at 2.5~\micron, and \citet{Moskovitz:2010} fit a 2$^{nd}$ order polynomial to the red-edge of Band II and take the value of the polynomial at 2.44~\micron.  In \S\ref{subsec:sara}, we define the red-edge as the reflectance value of a $5^{th}$ order polynomial fit to the entirety of Band II.  The choice is important to explore primarily for three reasons: 1) it potentially is a large systematic error to the Band II Area and BAR calculation;  2) it varies from study-to-study indicating that direct comparisons between band analysis studies should be done with caution; and 3) the Dunn 2010 Series, which provides calibration equations for S(IV) asteroids, is based on a Band II red edge set at 2.5~\micron.  The magnitude of this systematic error has only briefly explored for V-type asteroids with point (iv) defined at 2.4 and 2.5~\micron \ in \citet{Duffard:2006} with $\Delta BAR \approx 0.12 - 0.78$. 

For the S(IV) calibrations, this `Band II Red-Edge Problem' is particularly important because the Dunn 2010 Series red edge of 2.5~\micron \ is beyond the $\sim$2.45~\micron \ wavelength range where IRTF SpeX data is considered reliable.  This potentially introduces a systematic error in calculating Band II Centers, BARs, and the olivine-to-orthopyroxene ratio (Eqn.~\ref{eqn:ol_ratio}) inherent to all band parameter analysis studies performed with IRTF SpeX and other telescopic data with unreliable data beyond $\sim$2.45~\micron. %The calculated ol ratio is controlled by the BAR, such that larger BAR values translate to smaller ol ratios.  

\begin{figure}[htb]
\begin{center}
\includegraphics[width=12cm]{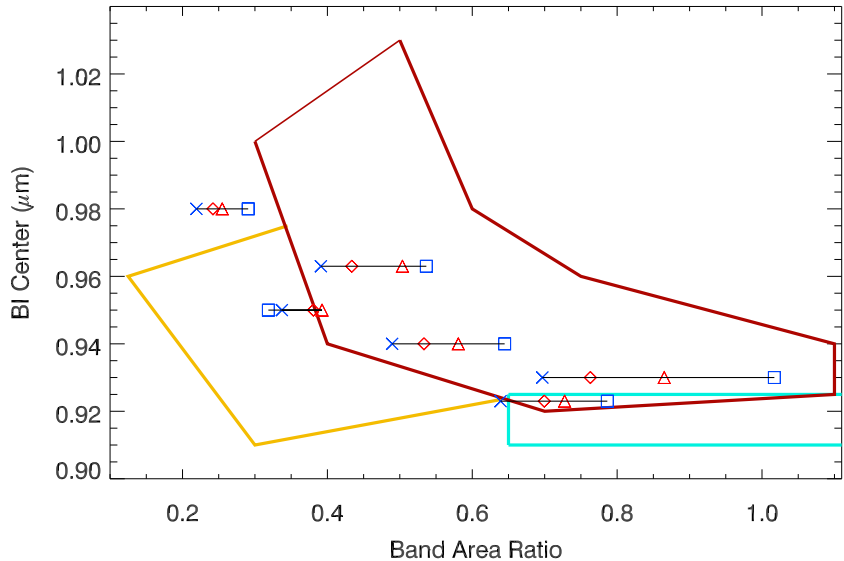}
\caption{The Band I Center versus BAR spectral parameters for the ordinary chondrites listed in Table~\ref{tab:relab_sample} over-plotted on the possible meteorite analog zones: OCs \citep[\emph{maroon,}][]{Gaffey:1993}, ureilite \citep[\emph{yellow,}][]{Cloutis:2010}, and primitive achondrite \citep[\emph{cyan,}][]{Burbine:2001}.  Four BAR calculations are displayed for a Band II red-edge set at 2.40 (\emph{X's}), 2.45 (\emph{diamonds}), 2.50 (\emph{triangles}), and 2.55~\micron \ (\emph{squares}) for each of the six OCs.  The 2.45 and 2.50~\micron \ red-edge BAR values are displayed in \emph{red} to highlight the difference between the Dunn 2010 Series red-edge value at 2.50, and the observationally limited red-edge value of 2.45~\micron.}
\label{fig:red_edge_bars}
\end{center}
\end{figure}

To test the magnitude of the `Red Edge Problem' error, we perform an investigation using six OC meteorite spectra from Brown University's Keck/NASA Reflectance Experiment Laboratory (RELAB) \citep{Pieters:2004}.  We chose two H, L, and LL chondrites (Table~\ref{tab:relab_sample}), and computed BARs using SARA for Band II red-edges set at 2.40, 2.45, 2.50, and 2.55~\micron.  Additionally, we computed BARs for two S(IV) MASs (1089 Tama and 1313 Berna) and two V-type asteroids (809 Lundia and 3703 Volkonskaya) using the same four red-edges.  These MASs were chosen because of their relatively high signal-to-noise longward of 2.45~\micron.  The BAR results for this investigation are shown in Table~\ref{tab:re_problem} and graphically represented in Fig.~\ref{fig:red_edge_bars}.

\begin{table} \scriptsize \renewcommand{\arraystretch}{0.8} \vspace{0.005in}
\caption{Meteorites measured in Red-Edge Investigation}
\begin{center}
\begin{tabular}{ p{2.3cm} p{0.5cm} p{2.0cm} p{2.0cm} } 

\toprule
Meteorite & Type & RELAB ID$^a$ & Grain size (\micron) \\ \toprule
Lost City & H & TB-TJM-129 & $<$150 \\
Schenectady & H & TB-TJM-083 & $<$150 \\
Karkh & L & TB-TJM-137 & $<$150 \\
Messina & L & TB-TJM-099 & $<$150 \\
Greenwell Springs & LL & TB-TJM-075 & $<$150 \\
%Karatu & LL & TB-TJM-077 & $<$75 \\
Bandong & LL & TB-TJM-067 & $<$150 \\ \bottomrule

\mbox{a Spectra are available on the RELAB database}
\mbox{~~ http://www.planetary.brown.edu/relabdata/}
%\mbox{(a) PDS Small Bodies Node, Binary Minor Planets V.6.0 \citep{Johnston:2013}}

\end{tabular}
\label{tab:relab_sample}
\end{center}
\end{table}
%If there is a poor telluric 

\begin{table} \scriptsize \renewcommand{\arraystretch}{0.8} \vspace{0.005in}
\caption{BAR as a function of Band II Red-Edge Choice}
\begin{center}
\begin{tabular}{ p{2.3cm} p{0.5cm} p{1.2cm} p{1.2cm} p{1.2cm} p{1.2cm} } 

\toprule
Meteorite & Type & 2.40 \micron & 2.45 \micron & 2.50 \micron & 2.55 \micron \\ \toprule
Lost City & H & 0.697 & 0.763 & 0.865 & 1.017 \\
Schenectady & H & 0.639 & 0.699 & 0.728 & 0.786 \\
Karkh & L & 0.337 & 0.381 & 0.393 & 0.318 \\
Messina & L & 0.489 & 0.533 & 0.581 & 0.645 \\
Greenwell Springs & LL & 0.391 & 0.434 & 0.504 & 0.537 \\
%Karatu & LL & 0.250 & 0.283 & 0.308 & 0.295 \\
Bandong & LL & 0.219 & 0.242 & 0.255 & 0.291 \\ \toprule
MAS & BDM &  &  &  &  \\ \toprule
809 Lundia & V & 1.838 & 1.999 & 2.113 & -- \\
3703 Volkonskaya & V & 1.771 & 1.919 & 1.961 & -- \\
1089 Tama & S & 0.602 & 0.623 & 0.624 & -- \\
1313 Berna & S & 0.365 & 0.460 & -- & -- \\  \bottomrule

%\mbox{Effective Diameter, $D_{eff}$ and Density citation:  \citep{Marchis:2012}}
%\mbox{$\dagger$ Wide binary system under \citet{Pravec:2007} nomenclature.}
%\mbox{(a) PDS Small Bodies Node, Binary Minor Planets V.6.0 \citep{Johnston:2013}}

\end{tabular}
\label{tab:re_problem}
\end{center}
\end{table}

There is a general trend that as the red-edge is extended to longer wavelengths, the BAR value increases.  The increase is due to more of Band II being added to the Band II area calculation, which is the numerator in the BAR calculation.  The one exception is the L chondrite, Karkh, which has a spectrum that exhibits a downturn in reflectance longward of 2.50~\micron, and thus SARA returns a lower Band II area and BAR value.  To evaluate the magnitude of the systematic error due to the `Red-Edge Problem', we consider the value $\Delta_{\rm BAR}$, which is the difference in BAR between 2.50~\micron \ and 2.45~\micron.  This value represents the BAR difference between the red-edge in the Dunn 2010 Series calibrations and the observationally limited red-edge in MAS spectral data. The $\Delta_{\rm BAR}$ values vary from meteorite to meteorite, with a smallest value of 0.012, and a largest value of 0.102.   In terms of calculated ol ratio, the largest $\Delta_{\rm BAR}$ translates to a potential error in ol ratio of 0.025.  This ol ratio error is slightly lower than the derived Dunn 2010 Series ol ratio error of 0.03.  This suggests that while the choice of red-edge is a potential source of error, the red-edge error itself is on the order of error inherent to the calibration, and therefore does not have a discernible influence on the calculated abundance ratios.  In other words, this conclusion supports the use of the Dunn 2010 Series calibration equations for red-edge choices that are $\gtrsim$~2.45~\micron.  For a red-edge choice at 2.4~\micron \ the difference in BAR value are larger, and in some cases exceed the error in the Dunn 2010 Series calibrations.  It should be noted that the increasing BAR value with increasing red-edge wavelength trend indicates the red-edge error is one-sided.  As such, the calculated ol ratio values for IRTF SpeX data are likely to be systematically larger by $\lesssim 0.025$ than the ol ratios listed in Table~\ref{tab:mineralogy}.

The influence of the red-edge problem for the MAS spectra is difficult to determine for the S-type asteroids.  This is mainly due to the low signal-to-noise longward of 2.45~\micron \ for IRTF SpeX data.  While SARA returns a BAR value for 1089 Tama with a 2.50~\micron \ red-edge, a visual inspection of the Band II definitions reveals that the 2.5~\micron \ reflectance value is lost in the noise; therefore, we do not trust the BAR values to be reflective of the data in this case.  The case is worse for 1313 Berna, where SARA is unable to return a valid Band II area.   For V-type MASs there is clear trend of increased BAR with increased red-edge wavelength position confirming the findings of \citet{Duffard:2006}.
%With respect to assigning a meteorite analog, however, the choice of red-edge is an important factor.  The typical spread of BAR values displayed in Fig.~\ref{fig:red_edge_bars} demonstrates

%This conclusion is reinforced when the $\Delta_{\rm BAR}$ for the S(IV) asteroids is considered.  The difference in BAR values between 2.45 and 2.50~\micron \ is much smaller for 10

%When considering the difference in BAR values between a 2.55~\micron \ and 2.40~\micron red-edge, the error increases to 0.078, with a median value near 0.035.
%If there is a poor telluric water feature near 1.9~\micron \ correction, then additional area may be added to Band II.  A larger area in Band II translates to a larger Band Area Ratio value shifting an asteroid to the right on Figs.~\ref{fig:gaffey_plot} and \ref{fig:gaffey_analogs}.  This potentially could lead to a misidentification of S-subtype, meteorite analog, as well as an underestimated ol/(ol + opx) ratio.  For example, consider the MAS 3309 Brorfelde (Fig.~\ref{fig:brorfelde}), which was observed under cloudy conditions and therefore has a poor telluric correction. 

%%%%%%%%%%%%%%%%%%%%%%%%%%%%%%%%%%%%%%%%%%%%%%%%%%%%%%
%%%%%%%%%%%%%%%%%%%%%%%%%%%%%%%%%%%%%%%%%%%%%%%%%%%%%%
\subsection{MAS formation considerations}
\label{sec:mas_formation}

As introduced in \S\ref{sec:macroporosity}, the porosity of a MAS is a measure of its internal structure.  As such, it provides a critical constraint to understanding formation and evolutionary pathways.  For example, consider the idealized MAS formation hypotheses through impact and ejection processes \citep{Durda:2004} and rotational effects acting on rubble-pile structures \citep{Pravec:2007,Descamps:2008,Jacobson:2011rot}.  Depending on the size and energy of an impactor, the MAS progenitor body can: 1) \emph{small impactor} - remain a `coherent' structure [porosity $\lesssim 15\%$] and eject material that could remain in orbit around the primary forming a T1 MAS \citep{Durda:2004}; 2) \emph{medium impactor} - fracture [porosity $\sim 15 - 30 \%$] and eject material that could remain in orbit around the primary forming a T1 MAS; or 3) \emph{large impactor} - catastrophically disrupt and reaggregate as a `rubble pile' [porosity $\gtrsim 30\%$] and eject material that could remain in orbit around the primary forming a T1 or T3 MAS \citep{Britt:2002, Pravec:2007,Descamps:2008, Jacobson:2011rot}.  A Rubble-pile can then evolve through rotational spin up through the Yarkovsky-O'Keefe-Radzievskii-Paddack (YORP) effect to form a MAS through mass-shedding (T1 or T3 MAS) or rotational fission forming a T4 MAS that may completely separate into a T2 MAS with nearly equally sized components \citep{Pravec:2006,Pravec:2007,Jacobson:2011rot}.  In the fission scenario, the entire systems can then be subsequently modified by tidal effects and the binary YORP (BYORP) effect, which can decay the orbits until the two components recombine forming a T4 MAS \citep{Pravec:2007,Descamps:2008,Jacobson:2011rot}.

%%%%%%%%%%%%%%%%%%%%%%%%%%%%%%%%%%%%%%%%%%%%%%%%%%%%%%
\subsubsection{T1 MASs}
\label{sec:t1mas}

Based upon the proposed formation hypotheses, the T1 MASs are the only systems that are not necessarily rubble-piles.  We find that the majority, but not all, of our T1s are rubble-pile structures.  The rubble-pile T1s are all large ($D_{\rm eff} > 140$ km) systems suggesting that the YORP effect is insufficient to increase their rotation rate to the point of mass-shedding for fission.  As such, we suggest that these T1 systems formed via impact and ejecta processes.  All of the T1 systems are also either C- or X-complex MASs with the exception of 107 Camilla, which we determine to be an L-type.

Interestingly, the two (121 Hermione and 130 Elektra) T1 MASs that exhibit a 0.7~\micron \ spectral feature, are also the only two coherent bodies in our sample.  The 0.7~\micron \ spectral feature is a likely indicator of aqueous alteration, which may be critical to understanding why both of these MASs, as well as taxonomically similar 93 Minerva \citep{Marchis:2011c}, have high densities compared to other C-type asteroids \citep{Marchis:2011c,Marchis:2012}.  The other C-types in the MAS sample are similar in size and semi-major axis suggesting that the presence of the 0.7~\micron \ feature may be linked to the internal porosity.   Possible explanations for this observation are: 1) the catastrophic disruption event required to create rubble-piles provides enough energy to dehydrate or sufficiently alter the aqueously altered minerals responsible for generating the 0.7~\micron \ feature; and/or 2) aqueous alteration is limited to the surfaces and the catastrophic disruption and re-aggregation even generates a new surface that spatially is relatively free of the original aqueous altered surface materials.   These explanations are not comprehensive or conclusive but they do warrant follow-up studies on MASs that exhibit a 0.7~\micron \ feature in an effort to understand the origins, timing, and evolution of this feature and aqueous alteration in the asteroid belt.  If borne out, they suggest that the presence of the 0.7~\micron \ feature is linked with the collisional history of the asteroid.  Consistent with this suggestion is the lower fraction of Themis and Hygiea family members that exhibit a 0.7~\micron \ feature \citep{Rivkin:2012}.  As dynamical family members, it is likely that these asteroids are the result of a catastrophic disruption event.
% The connection between aqueous alteration and lower macroporosity suggests that their may be a link between how these systems formed and aqueous alteration. 

%%%%%%%%%%%%%%%%%%%%%%%%%%%%%%%%%%%%%%%%%%%%%%%%%%%%%%
\subsubsection{Rubble-pile types: T2, T3, T4, and W MASs}
\label{sec:rubble_mas}

%General
All of the small ($D_{\rm eff} \lesssim$ 15 km) MASs are consistent within errors with rubble-pile structures.  These MASs also all belong to either the T2, T3, or the \citet{Pravec:2007} W group.  These MASs being rubble-piles is consistent with the hypothesis that T2 and T3 MASs are products of the evolutionary track of a rubble-pile asteroid \citep{Jacobson:2011rot}. The formation of the W group, which includes well separated, small binary systems, is not well understood, but the potential rubble-pile structure of 3749 Balam suggests that the W group is also a product of rubble-pile physics.  Indeed, \citet{Jacobson:2014} find that 3749 Balam and 4674 Pauling are consistent with a recently proposed formation hypotheses to explain the formation of wide asynchronous systems that requires rotational fission of a rubble-pile.  We expect that a macroporosity determination of 4674 Pauling will reveal that it also has a rubble-pile structure.

%T2s
The expectation is that T2 MASs (large or small) are rubble-pile structures formed via rotational fission processes, and indeed 809 Lundia, 1089 Tama, and 1313 Berna fall within the rubble-pile macroporosity range within their uncertainties.  1089 Tama falls near the boundary between fractured and rubble-pile making a possible case for formation via a process other than rotational fission.  However, if the condition for `rubble-pile' is moderately relaxed to $P_m \gtrsim 25\%$ then the likelihood of 1089 Tama being a fractured body diminishes considerably.  Such a relaxation of the boundary is not unfounded as the internal structure divisions described in \citep{Britt:2002} are not well constrained.

%Our sample of MASs with estimated macroporosity also includes 3749 Balam, which is a wide, asynchronous binary \citep{Pravec:2007,Jacobson:2014}.
%T2 via rotational fission is also observationally supported as all T2 systems spread along the same equilibrium sequence of two fluids approximated by two identical ellipsoids rotating around their center of gravity \citep{Marchis:2012}.  Unexpectedly, however, T2 MAS 1089 Tama ($P_m = 21.18 \pm 6.25$ for L-chondrite analog) falls within the fractured category.  If the condition for `rubble-pile' is moderately relaxed to $P_m \gtrsim 25\%$, as has been suggested by some researchers [ref], then, within errors, there is no violation of the formation via rotational fission hypothesis \citep{Descamps:2008}.  [Can we say anything about whether 1089 is a novel system?  If equal sized components, then nothing special... different sized components?  that would be novel.]

%%%%%%%%%%%%%%%%%%%%%%%%%%%%%%%%%%%%%%%%%%%%%%%%%%%%%%
%\subsection{Individual objects}
%\label{sec:rubble_mas}

%%%%%%%%%%%%%%%%%%%%%%%%%%%%%%%%%%%%%%%%%%%%%%%%%%%%%%
%%%%%%%%%%%%%%%%%%%%%%%%%%%%%%%%%%%%%%%%%%%%%%%%%%%%%%
\subsection{LL chondrite MAS mineralogies connection to Flora Family}
\label{sec:too_many_LLs}

Among the 15 MASs with OC-like composition, we find that most fall within the mineralogical bounds of the LL chondrites (Fig.~\ref{fig:fs}) with an additional MAS that is on the L/LL boundary.   If we include the L/LL boundary object with the LL-like MASs, we have 13.3\%, 20.0\%, and 66.7\% of our OC-like MASs being H-, L-, and LL-like, respectively.  The observed excess of LL chondrite mineralogies is surprising given the percentages of H, L, and LL OC meteorite falls on Earth, which have a break down of 42.6\%, 47.5\%, and 9.8\% \citep{Burbine:2002}.  Even more intriguing is a similar excess of LL chondrite mineralogies is observed for the near-Earth asteroid (NEA) population \citep{Vernazza:2008,deLeon:2010,Dunn:2013nea,Thomas:2014}.  Based on dynamical models, an effective delivery system for the kilometer-sized NEA population is the $\nu_6$ secular resonance with Saturn located at the innermost edge of the main belt \citep{Bottke:2002}.  Located near this escape-hatch is the Flora family, which is expected to contribute significantly to the NEA population \citep{Nesvorny:2002}.  As such, the Flora family is often considered to be the source of the excess of LL-like mineralogies observed in the NEA population.
%Based upon dynamical simulations and the proximity to the $\nu_6$ secular resonance with Saturn, these researchers have suggested that the Flora family is the source of the LL chondrite excess in the NEA population. % While the cause of the discrepancy between number of falls and asteroids that are spectrally analogous to LL chondrites remains unclear, our MAS spectral survey supports the connection between the main belt Flora family and the NEA population.  

Of the ten MASs that exhibit LL chondrite mineralogies, seven fall within the narrow semi-major axis range of 2.17 - 2.25 AU; eccentricity range of 0.11 - 0.19; and inclination range of 3 - 6$^\circ$.  This orbital space is consistent with the Flora dynamical family, and indeed, four of the seven MASs are currently identified as part of the Flora family.   Our excess of LL-like mineralogies (ol ratio = 56--72\%, Dunn 2010 Series) being associated with the Flora family is in agreement with the Flora family having olivine-rich mineralogies \citep{Vernazza:2008}.  As such, it provides additional evidence for the Flora family being a source for the overabundance of LL-like mineralogies in the NEA population.
%However, this is the first reporting of the Flora family being consistent with LL-like mineralogies, and it supports the hypothesis that the Flora family is indeed the source of LL chondrites.

Additionally, the steep size distribution of the Flora family suggests a collisional breakup origin \citep{Cellino:1991}.  In support of this are the seven MASs consistent with the Flora family membership.  Of these seven MASs, all are small ($D_{\rm eff} < 15$ km); and one is a T2 MAS, five are T3 MASs, and one is unclassified.  All of these systems are likely to have formed via rubble-pile physics \citep{Jacobson:2011rot,Jacobson:2014}.  This is consistent with the \citet{Cellino:1991} suggestion that the origin of the Flora family is the catastrophic disruption of a progenitor body.  The consistency of LL-like mineralogies within this group suggests that the progenitor body is the likely source for some proportion of the LL chondrites. 
%This suggests that many of the members being of rubble-pile structure that potentially could evolve into T2 or T3 MAS systems following the rubble-pile physics evolution pathways of \citet{Jacobson:2011rot}.  This well-matches the MAS type distribution for the observed objects with LL chondrite mineralogies: one T2 and nine T3 systems.   

Two MASs with LL chondrite mineralogies, 1313 Berna and 1333 Cevenola, are also closely clustered with semi-major axes at 2.66 and 2.63 AU, respectively, and have eccentricities and inclinations consistent with the Eunomia family.  The Eunomia family is situated near the 3:1 and 8:3 mean motion resonances with Jupiter that provide a means to deliver asteroids to the NEA population and meteorites to Earth \citep{Morbidelli:1995,Bottke:2002}.  Albeit small numbers, these two MASs suggest that the Eunomia family may be another source of LL chondrites indicating that the LL chondrites may have more than one associated parent body.

%%%%%%%%%%%%%%%%%%%%%%%%%%%%%%%%%%%%%%%%%%%%%%%%%%%%%%
%%%%%%%%%%%%%%%%%%%%%%%%%%%%%%%%%%%%%%%%%%%%%%%%%%%%%%
\section{Summary}
\label{sec:conclusion}  

We have performed a large-scale observing and spectral analysis study on 42 MASs.  The study covers both visible (0.45 - 0.85~\micron) and NIR (0.75 - 2.45~\micron) spectral ranges, and the all of the data are processed in a consistent way to make the results comparable.   When available, our spectral data are supplemented using publicly available data available through SMASSII, S3OS2, and PDS Small Bodies Node.  The processed MAS data set allows us to classify 21 new Bus-DeMeo taxonomies using  MIT's Bus-DeMeo Taxonomy Classification Tool.

Our MAS spectral data set is further analyzed to determine mineralogy and meteorite analogs. This analysis is performed using the new, IDL-based, Spectral Analysis Routine for Asteroids (SARA).   SARA is a generalized band parameter analysis algorithm that  is currently configured for S-complex asteroids and V-type asteroids.  To make our calculated band parameters comparable to laboratory and calibration studies, the Band II centers, areas, and BARs are corrected for temperature effects using the equations of \citet{Sanchez:2012}.  The final corrected band parameters are then employed to determine the S-subtypes \citep{Gaffey:1993}, relative and modal abundances, and meteorite analogs for the MASs.  The determined meteorite analogs include OCs, basaltic achondrites, olivine-rich meteorites, and potentially primitive achondrites.  For the MASs with OC (S[IV]s) and basaltic achondrite (V-types) mineralogies, we compare the calculated band parameters and derived mineralogy to determine meteorite analogs to the subgroup level, i.e., H, L, or LL chondrite for OCs; and Howardite, Eucrite, or Diogenite for basaltic achondrites.  For the MASs with OC mineralogies, we find an excess of LL chondritic material (10/16) and few MASs consistent with L (3/16) or H (2/16) material.  One object, 2131 Mayall, does not have a currently identifiable analog, but may be associated with the primitive achondrites (lodranites or acapulcoites) \citep{Burbine:2001,Lucas:2014}.

The meteorite analogs are applied to extend the MAS density results of \citet{Marchis:2012} to estimate macroporosity.  The determined macroporosities allow us to determine the internal structure of the MASs, which is applied to evaluate the formation hypotheses for MASs.  Of the thirteen macroporosity estimates we were able to determine, the majority are consistent with rubble-piles with only two objects being consistent with coherent bodies.  The small MASs ($D_{\rm eff} \lesssim $15 km) are all found to be rubble-piles and either T2 or T3 MASs, consistent with formation and evolution via rubble-pile physics \citep{Jacobson:2011rot,Jacobson:2014}.  The two coherent bodies, 121 Hermione and 130 Elektra, are both large ($D > 100$ km ), T1 MASs, which is consistent with the small satellites forming via ejection of material off the surface following a impact \citep{Durda:2004}.  
%In this visible and NIR spectral observing campaigns targeting main belt MAS, we have observed and analyzed a 42 MASs.   This is the largest, self-consistent, spectral survey of MASs to date

In addition to this summary, we extract several conclusions and suggestions from the analysis of the MAS spectral data:

\begin{enumerate}

\item{The choice of the Band II red-edge is a source of error when calculating the BAR, and it makes direct BAR comparisons between studies that use a different red-edge difficult.  However, the magnitude of the `Red-edge Problem' error when applying the mineralogy calibrations in the Dunn 2010 Series is smaller than the error in the calibrations equations. Therefore, the `Red-edge Problem' is not statistically discernible in the mineralogy.  On the other hand, the red-edge error is one-sided, and thus systematic in that the choice of a red-edge at 2.45~\micron \ compared to 2.50~\micron \ will offset in olivine-to-orthopyroxene ratio to larger values by $\lesssim$ 0.025.}

\item{The two MASs we find to be consistent with coherent bodies are the only C-type asteroids that exhibit the 0.7~\micron \ feature that is indicative of aqueous alteration.  This is suggestive, but not conclusive, of a causal link between the presence of the 0.7~\micron \ feature and the collisional history of the body it is observed on.  However, follow-up spectroscopic studies of MASs with a 0.7~\micron \ feature are required to test the validity of this suggestion.}

\item{The overabundance of LL-like mineralogies in our S(IV) sample is explained by Flora family membership. Seven out of the ten MASs with OC-like mineralogies fall within the orbital parameter description of the Flora family, and all seven are analogous to the LL chondrites.  This overabundance is consistent with studies of NEAs, which find a similar excess of LL-like mineralogies, strengthening the proposed connection between LL chondrites and the Flora family.  These systems are also rubble-piles consistent with the Flora family forming via the catastrophic disruption suggesting a progenitor body with an LL-like mineralogy.  Moreover, we note that two LL-like MASs consistent with the Eunomia family are also both LL-like suggesting the Eunomia family as a possible additional source for LL-like NEAs and LL chondrites on Earth.}

\end{enumerate}

%%%%%%%%%%%%%%%%%%%%%%%%%%%%%%%%%%%%%%%%%%%%%%%%%%%%%%
%%%%%%%%%%%%%%%%%%%%%%%%%%%%%%%%%%%%%%%%%%%%%%%%%%%%%%
\section{Acknowledgements}

The authors would like to express their appreciation and acknowledge several individuals who have generous to aid in this publication.  Thank you to Vishnu Reddy and Juan Sanchez for providing their spectrum to 5407 (1992 AX) for comparison to our own spectra.  Thank you to Paul Hardersen and Michael Lucas for acting as a beta testers for the SARA algorithm.  Finally, thank you to Tasha Dunn and the anonymous reviewer for their thoughtful comments that have much improved the quality of this article.
We express our appreciation to agencies that provided funding for this research.  Specifically, we thank the NASA Planetary Astronomy (PAST) grant program (NNX 11AD62G) for their support.  Marcelo Assafin thanks the CNPq (grants 473002/2013-2 and 308721/2011-0) and FAPERJ (grant E-26/111.488/2013).
We would also like to acknowledge the RELAB database and PDS Small Bodies Node data ferret as an excellent resource to the planetary astronomy community.  

%%%%%%%%%%%%%%%%%%%%%%%%%%%%%%%%%%%%%%%%%%%%%%%%%%%%%%
\appendix

\setcounter{figure}{0}
\renewcommand{\thefigure}{A\arabic{figure}}

\begin{sidewaysfigure}[htb]
	\includegraphics[height=6.5in,width=8.25in]{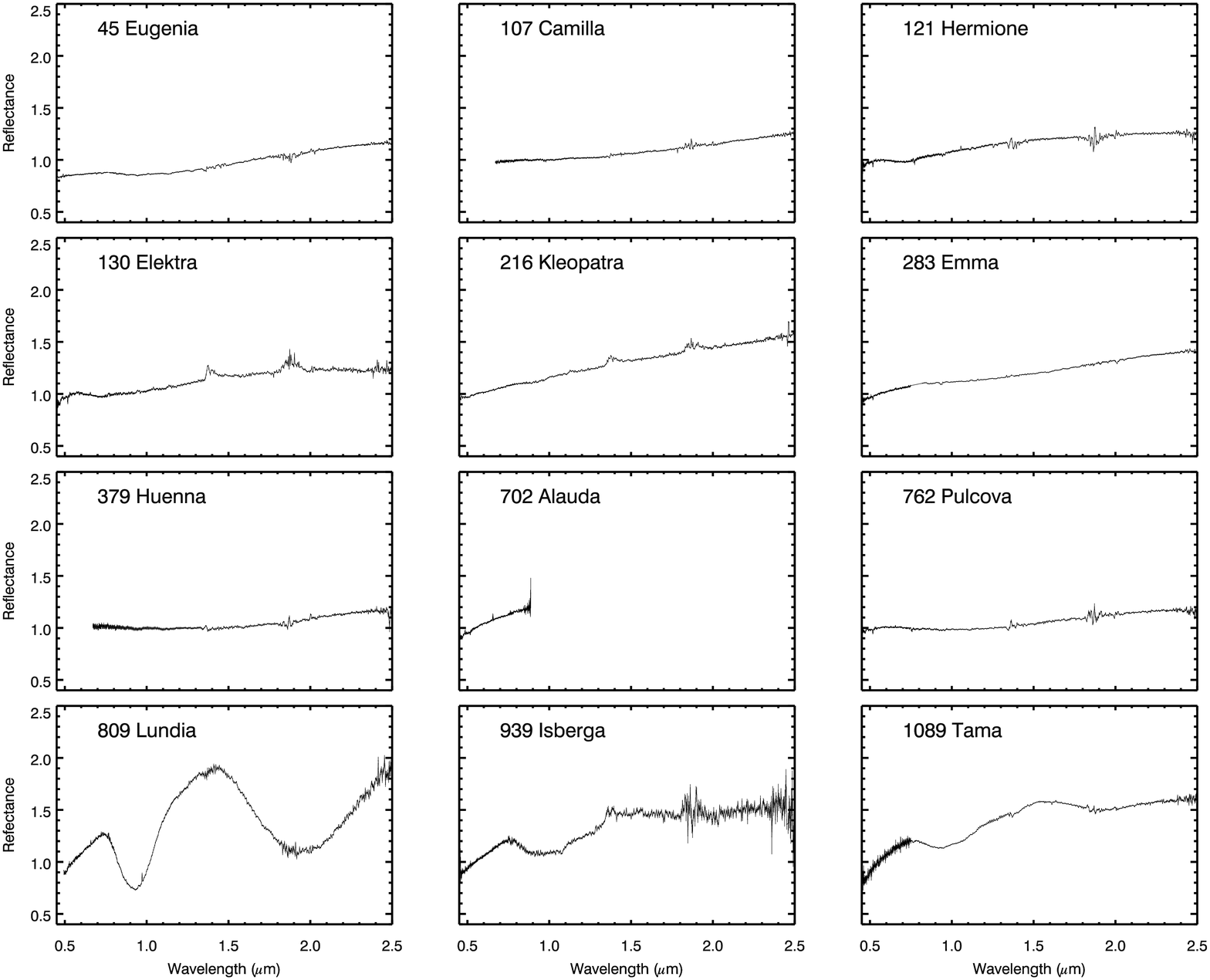} 
	\caption{The visible and near-infrared spectra of 42 main belt MASs used in this study.  All spectra with visible data are normalized to unity at 0.55~\micron.  If only infrared spectral data are available, then the spectra are normalized to unity at 1.0 \micron.}
	\label{fig:spectra_thumbs}
\end{sidewaysfigure}
	
\begin{sidewaysfigure}[htb]
	\includegraphics[height=6.5in,width=8.25in]{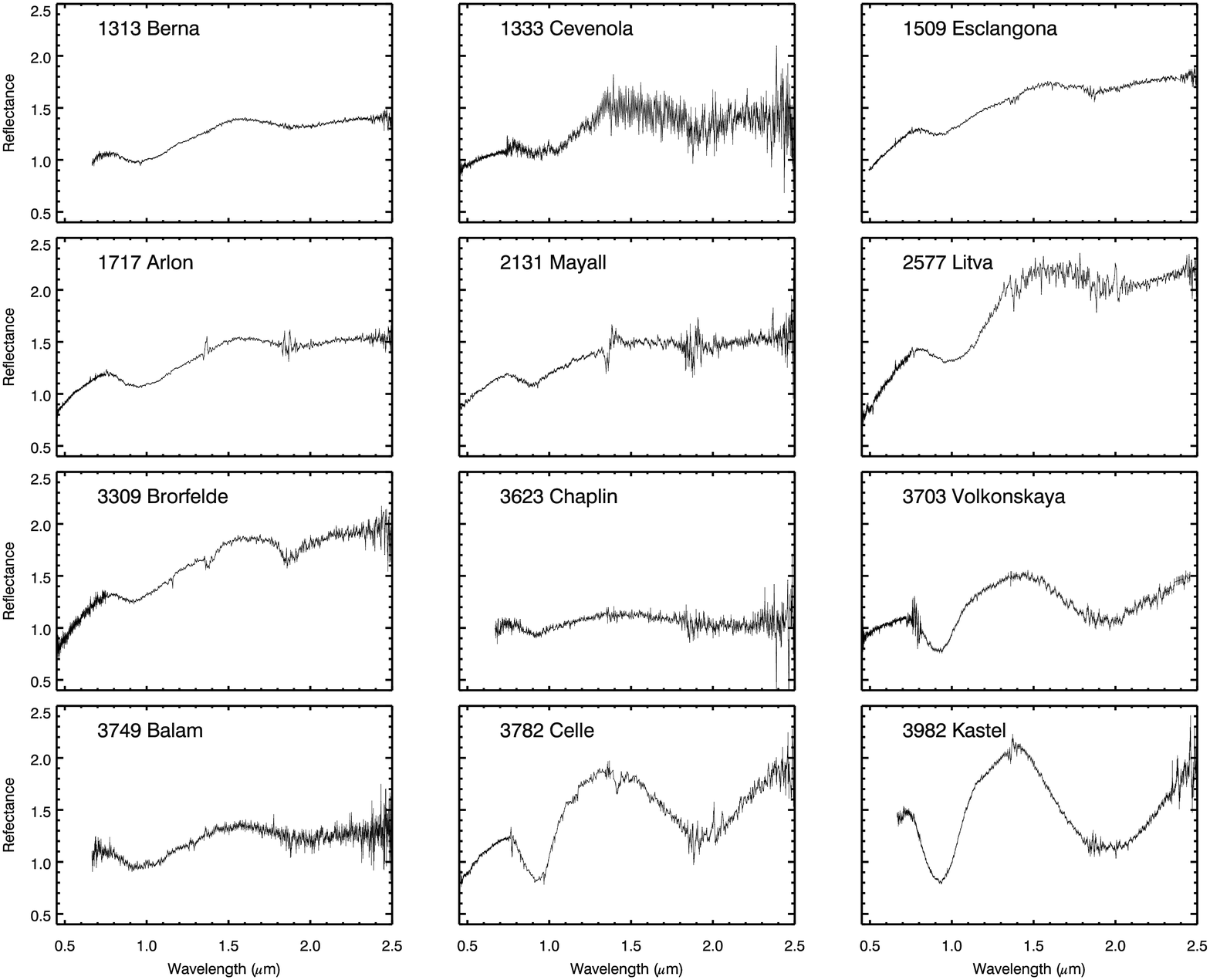} 
	\caption{\emph{continued}}
\end{sidewaysfigure}

\begin{sidewaysfigure}[htb]
	\includegraphics[height=6.5in,width=8.25in]{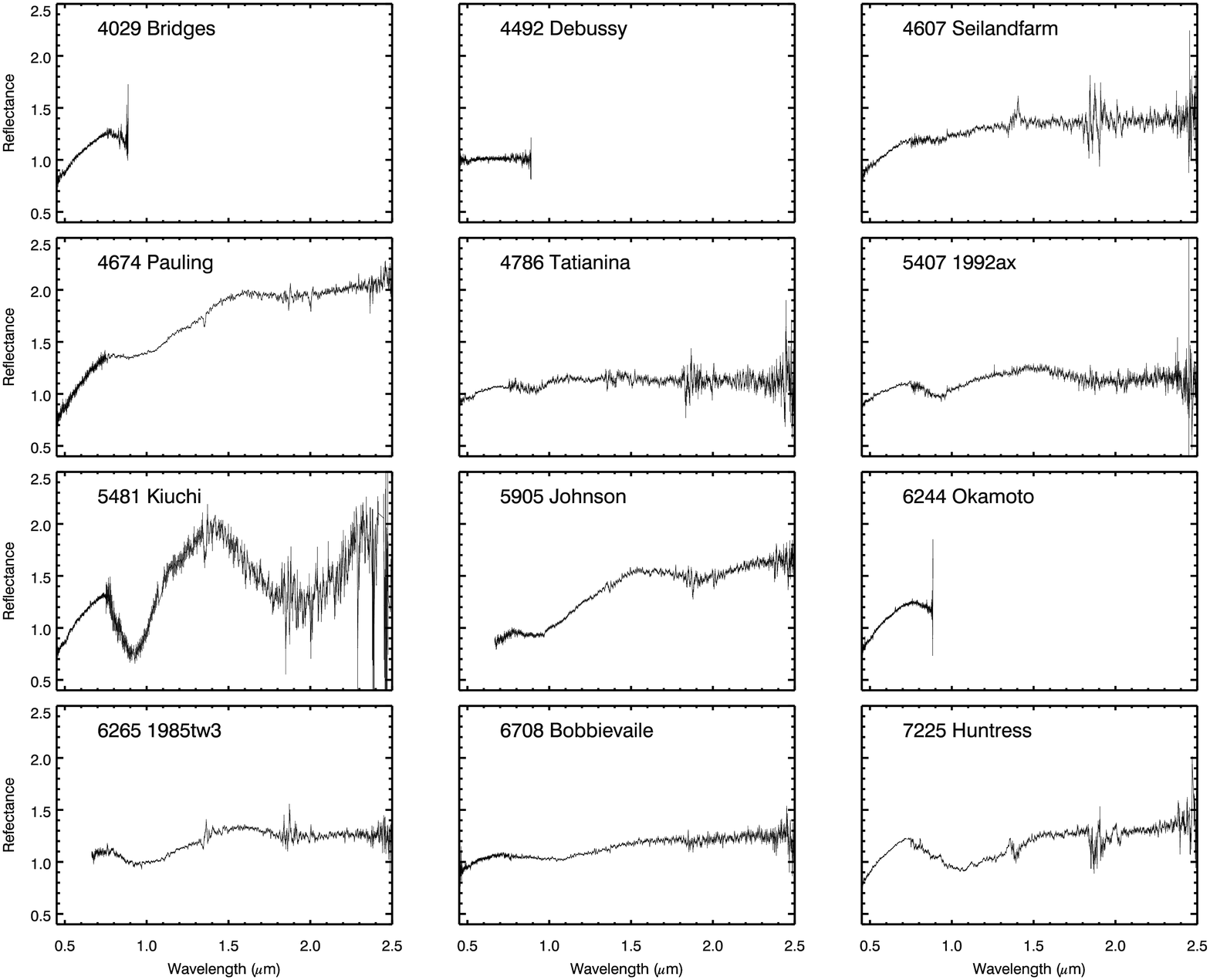} 
	\caption{\emph{continued}}
\end{sidewaysfigure}

\begin{sidewaysfigure}[htb]
	\includegraphics[height=6.5in,width=8.25in]{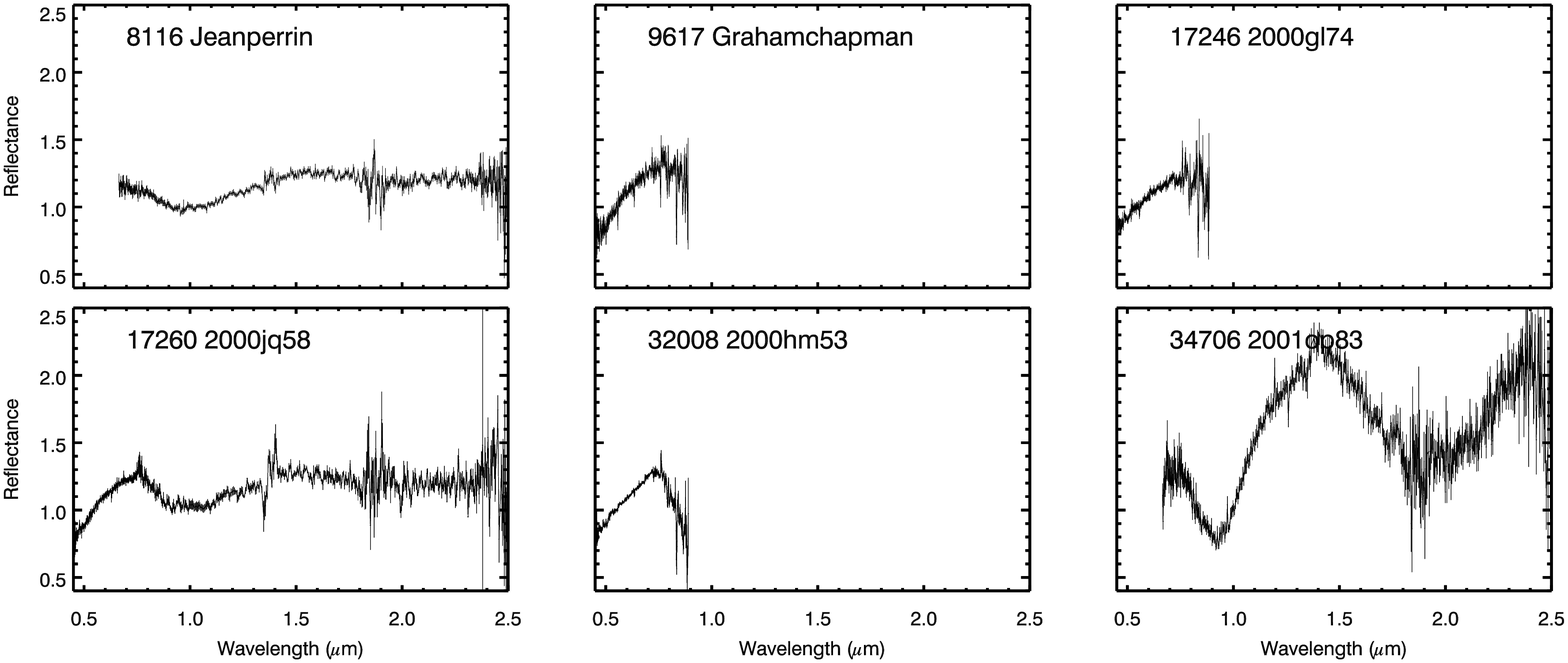}
	\caption{\emph{continued}}
\end{sidewaysfigure}

%%%%%%%%%%%%%%%%%%%%%%%%%%%%%%%%%%%%%%%%%%%%%%%%%%%%%%
\bibliographystyle{elsarticle-harv}
\bibliography{asteroids}

\end{document}